\documentclass[12pt]{article}
\usepackage{amsmath,amssymb}
\usepackage{graphicx}
\newcommand{\eqdef}{\stackrel{\text{def}}{=}}
\newcommand{\eqdefrm}{\stackrel{\text{\rm def}}{=}}
\newcommand{\n}{\nonumber\\}
\newcommand{\bm}{\boldsymbol}
\newcommand{\ignore}[1]{}
\numberwithin{equation}{section}
\newcommand{\Romannumeral}[1]{\uppercase\expandafter{\romannumeral#1}}
\newcommand{\I}{\text{\Romannumeral{1}}}
\newcommand{\II}{\text{\Romannumeral{2}}}

\newtheorem{prop}{\bf Proposition}
\newtheorem{thm}{\bf Theorem}

\newcounter{myremarkbangou}
\renewcommand{\themyremarkbangou}{\arabic{myremarkbangou}}
\newcommand{\remark}{\refstepcounter{myremarkbangou}
\noindent{\bf Remark \themyremarkbangou\ }}
\newcounter{myremarkbangouA}

\newcommand{\cH}{\mathcal{H}}
\newcommand{\wtcH}{\widetilde{\mathcal{H}}}
\newcommand{\cA}{\mathcal{A}}
\newcommand{\cE}{\mathcal{E}}
\newcommand{\cF}{\mathcal{F}}
\newcommand{\cB}{\mathcal{B}}
\newcommand{\cD}{\mathcal{D}}
\newcommand{\constF}{c_{\text{\tiny$\mathcal{F}$}}}
\newcommand{\heihoukon}{\raisebox{1mm}{$\sqrt{~~~}\,$}}

\allowdisplaybreaks[4]

\begin{document}

\baselineskip=20pt

\newcommand{\preprint}{
\vspace*{-20mm}
   \begin{flushright}\normalsize \sf
    DPSU-25-2\\
  \end{flushright}}
\newcommand{\Title}[1]{{\baselineskip=26pt
  \begin{center} \Large \bf #1 \\ \ \\ \end{center}}}
\newcommand{\Author}{\begin{center}
  \large \bf Satoru Odake \end{center}}
\newcommand{\Address}{\begin{center}
     Faculty of Science, Shinshu University,
     Matsumoto 390-8621, Japan
   \end{center}}
\newcommand{\Accepted}[1]{\begin{center}
  {\large \sf #1}\\ \vspace{1mm}{\small \sf Accepted for Publication}
  \end{center}}

\preprint
\thispagestyle{empty}

\Title{Some Difference Relations for Orthogonal Polynomials of a Continuous
Variable in the Askey Scheme}

\Author

\Address
\vspace{1cm}

\begin{abstract}
Orthogonal polynomials of a continuous variable in the Askey scheme satisfying
second order difference equations, such as the Askey-Wilson polynomial, can be
studied by the quantum mechanical formulation, idQM (discrete quantum mechanics
with pure imaginary shifts). These idQM systems have the shape invariance
property, which relates the Hilbert space $\mathsf{H}_{\bm{\lambda}}$
($\bm{\lambda}$ : a set of parameters) and that with shifted parameters
$\mathsf{H}_{\bm{\lambda}+\bm{\delta}}$ ($\bm{\delta}$ : shift of
$\bm{\lambda}$), and gives the forward and backward shift relations for the
orthogonal polynomials.
Based on the forward shift relation and the Christoffel's theorem with some
polynomial $\check{\Phi}(x)$, which is expressed in terms of the quantities
appeared in the forward and backward shift relations,
we obtain some difference relations for the orthogonal polynomials.
The multiplication of $\sqrt{\check{\Phi}(x)}$ gives a surjective map from
$\mathsf{H}_{\bm{\lambda}+2\bm{\delta}}$ to $\mathsf{H}_{\bm{\lambda}}$.
Similarly, for the orthogonal polynomials in the Askey scheme satisfying
second order differential equations, such as the Jacobi polynomial, we obtain
some differential relations, and the multiplication of
$\sqrt{\check{\Phi}(x)}$ in this case gives a surjective map from
$\mathsf{H}_{\bm{\lambda}+\bm{\delta}}$ to $\mathsf{H}_{\bm{\lambda}}$.
\end{abstract}

\section{Introduction}
\label{sec:intro}

The Askey scheme of (basic) hypergeometric orthogonal polynomials is
a classification of the orthogonal polynomials satisfying second order
differential or difference equations allowed by the Bochner's theorem and
its generalizations \cite{ismail,kls}.
These polynomials can be studied by the quantum mechanical formulations:
ordinary quantum mechanics (oQM) and two kinds of discrete quantum mechanics
(dQM), dQM with pure imaginary shifts (idQM) and dQM with real shifts (rdQM)
\cite{os24}. The Schr\"odinger equation for oQM is a (second order)
differential equation and that for dQM is a (second order) difference equation.
The coordinate $x$ for oQM and idQM is continuous and that for rdQM is discrete.

The eigenfunctions of the quantum systems associated with the orthogonal
polynomials in the Askey scheme, $\phi_n(x)$ ($n\in\mathbb{Z}_{\geq0}$)
($n\in\{0,1,\ldots,N\}$ for finite rdQM systems), have the following forms
\cite{os24}:
\begin{equation}
  \phi_n(x)=\phi_0(x)\check{P}_n(x),\quad
  \check{P}_n(x)\eqdef P_n\bigl(\eta(x)\bigr).
\end{equation}
Here $\phi_0(x)$ is the ground state wavefunction, $\eta(x)$ is the
sinusoidal coordinate and $P_n(\eta)$ is the orthogonal polynomial of degree
$n$ in $\eta$ with respect to the weight function $\phi_0(x)^2$.
These quantum mechanical systems have the shape invariance property,
$\cA(\bm{\lambda})\cA(\bm{\lambda})^{\dagger}
=(\text{constant})\times\cA(\bm{\lambda}+\bm{\delta})^{\dagger}
\cA(\bm{\lambda}+\bm{\delta})+(\text{constant})$,
where $\cA(\bm{\lambda})$ is some operator and
$\bm{\lambda}=(\lambda_1,\lambda_2,\ldots)$ is a set of parameters contained
in the system and $\bm{\delta}$ is its shift.
This is a sufficient condition for the solvability.
The shape invariance combined with the Crum's theorem and its generalizations
\cite{crum,os15,os22} gives the relation between the Hilbert space
$\mathsf{H}_{\bm{\lambda}}$ and that with shifted parameters
$\mathsf{H}_{\bm{\lambda}+\bm{\delta}}$, and gives the forward and backward
shift relations for the orthogonal polynomials.

Recently, Ismail and Saad studied the Meixner-Pollaczek polynomial
$P^{(a)}_n(x;\phi)$ (change of notation : $\lambda\to a$) and obtained various
relations \cite{is25}.
We consider eq.(21) in \cite{is25}, which is a relation for
$P^{(a+1)}_n(x;\phi)$ and $P^{(a)}_{n+k}(x;\phi)$ ($k=0,1,2$), derived from
the Christoffel's theorem (Theorem 2.7.1 in \cite{ismail}).
In idQM formulation ($\bm{\lambda}=(a,\phi)$, $\bm{\delta}=(\frac12,0)$,
$\eta(x)=x$,
$\phi_0(x;\bm{\lambda})^2=e^{(2\phi-\pi)x}\Gamma(a+ix)\Gamma(a-ix)$),
the ratio of $\phi_0(x;\bm{\lambda}+2\bm{\delta})^2$ and
$\phi_0(x;\bm{\lambda})^2$ is a polynomial in $\eta(x)$, and the
Christoffel's theorem can be applied.
This fact (the ratio of $\phi_0(x;\bm{\lambda}+2\bm{\delta})^2$ and
$\phi_0(x;\bm{\lambda})^2$ is a polynomial in $\eta(x)$) holds for all idQM
systems associated with the orthogonal polynomials in the Askey scheme, and
this ratio is expressed in terms of the quantities appeared in the forward
and backward shift relations.
{}From the Christoffel's theorem, we can write down a relation for
$\check{P}_n(x;\bm{\lambda}+2\bm{\delta})$ and
$\check{P}_{n+k}(x;\bm{\lambda})$ explicitly.
Moreover, by combining this relation with the forward shift relation, we obtain
some difference relation for $\check{P}_n(x;\bm{\lambda})$.
Similar calculation can be applied to oQM systems. But in this case, the
ratio of $\phi_0(x;\bm{\lambda}+\bm{\delta})^2$ and
$\phi_0(x;\bm{\lambda})^2$ is a polynomial in $\eta(x)$.

This paper is organized as follows.
In section \ref{sec:QMF}, we recapitulate the quantum mechanical formulation,
idQM.
In section \ref{sec:CT}, we recall the Christoffel's theorem.
Section \ref{sec:DR} is a main part of this paper.
We write down the Christoffel's theorem and obtain some difference relations
for the orthogonal polynomials in idQM explicitly.
Their explicit forms are given in \S\,\ref{sec:ExAW} for the Askey-Wilson
polynomial and in Appendix \ref{app:ExidQM} for other polynomials.
We also provide other properties.
In section \ref{sec:DlR}, some differential relations for the orthogonal
polynomials in oQM are obtained explicitly.
Their explicit forms are given in \S\,\ref{sec:ExJ} for the Jacobi
polynomial and in Appendix \ref{app:ExoQM} for other polynomials.
Section \ref{sec:summary} is for a summary and comments.

\section{Quantum Mechanical Formulation}
\label{sec:QMF}

In this section we recapitulate the quantum mechanical formulation:
the discrete quantum mechanics with pure imaginary shifts (idQM)
\cite{os13,os24}.
Explicit forms of various quantities are given in \S\,\ref{sec:ExAW} and
Appendix \ref{app:ExidQM}.

The dynamical variables of idQM are the real coordinate $x$
($x_{\text{min}}<x<x_{\text{max}}$)
and its conjugate momentum $p=-i\partial_x$, which are governed by the
following factorized positive semi-definite Hamiltonian:
\begin{align}
  &\cH\eqdef\sqrt{V(x)}\,e^{\gamma p}\sqrt{V^*(x)}
  +\!\sqrt{V^*(x)}\,e^{-\gamma p}\sqrt{V(x)}
  -V(x)-V^*(x)=\cA^{\dagger}\cA,
  \label{H}\\
  &\cA\eqdef i\bigl(e^{\frac{\gamma}{2}p}\sqrt{V^*(x)}
  -e^{-\frac{\gamma}{2}p}\sqrt{V(x)}\,\bigr),\quad
  \cA^{\dagger}=-i\bigl(\sqrt{V(x)}\,e^{\frac{\gamma}{2}p}
  -\sqrt{V^*(x)}\,e^{-\frac{\gamma}{2}p}\bigr).
\end{align}
Here the potential function $V(x)$ is an analytic function of $x$ and $\gamma$
is a real constant.
The $*$-operation on an analytic function $f(x)=\sum_na_nx^n$
($a_n\in\mathbb{C}$) is defined by $f^*(x)=\sum_na_n^*x^n$, in which
$a_n^*$ is the complex conjugation of $a_n$.
The symbol $\heihoukon$ is the square root as a complex function, with the
appropriate branch chosen.
An exponential function of $p$, $e^{\alpha p}$ ($\alpha$ : constant), is a
shift operator $e^{\alpha p}f(x)=f(x-i\alpha)$.
The Schr\"{o}dinger equation $\cH\phi(x)=\cE\phi(x)$ is an analytic difference
equation with pure imaginary shifts.
The inner product of two wave functions $\phi(x)$ and $\psi(x)$ is given by
$(\phi,\psi)\eqdef\int_{x_{\text{min}}}^{x_{\text{max}}}dx\,\phi^*(x)\psi(x)$
and the norm of a wave function $\phi(x)$ is
$|\!|\phi|\!|\eqdef\sqrt{(\phi,\phi)}$.

The idQM systems corresponding to the orthogonal polynomials in the Askey
scheme have an infinite number of square-integrable eigenstates $\phi_n(x)$
with discrete energy levels $\cE_n$:
\begin{equation}
  \cH\phi_n(x)=\cE_n\phi_n(x)\ \ (n\in\mathbb{Z}_{\geq 0}),
  \quad 0=\cE_0<\cE_1<\cE_2<\cdots.
  \label{Hphi=}
\end{equation}
The eigenfunctions $\phi_n(x)$ can be chosen `real', $\phi_n^*(x)=\phi_n(x)$,
and the groundstate wavefunction $\phi_0(x)$ is determined as the zero mode of
the operator $\cA$, $\cA\phi_0(x)=0$.
The eigenfunctions $\phi_n(x)$ have the following factorized form:
\begin{equation}
  \phi_n(x)=\phi_0(x)\check{P}_n(x),\quad
  \check{P}_n(x)\eqdef P_n\bigl(\eta(x)\bigr)\ \ (n\in\mathbb{Z}_{\geq 0}),
  \label{phin=}
\end{equation}
where $\eta=\eta(x)$ is the sinusoidal coordinate and $P_n(\eta)$ is a
polynomial of degree $n$ in $\eta$ and $P_{-1}(\eta)\eqdef 0$.
The second order difference operator $\wtcH$ acting on the polynomial
eigenfunctions is square root free:
\begin{align}
  &\wtcH\eqdef\phi_0(x)^{-1}\circ\cH\circ\phi_0(x),
  \label{wtcHdef}\\
  &\rightarrow\wtcH=V(x)(e^{\gamma p}-1)+V^*(x)(e^{-\gamma p}-1),
  \label{wtcH}\\
  &\wtcH\check{P}_n(x)=\cE_n\check{P}_n(x)\ \ (n\in\mathbb{Z}_{\geq 0}).
\end{align}
Since the Hamiltonian $\cH$ is a hermitian operator, we have the orthogonality
relations,
\begin{equation}
  (\phi_n,\phi_m)=h_n\delta_{nm}
  \ \ (n,m\in\mathbb{Z}_{\geq 0}),\quad 0<h_n<\infty.
\end{equation}
This gives the orthogonality relations for the polynomials,
\begin{equation}
  \int_{x_{\text{min}}}^{x_{\text{max}}}dx\,\phi_0(x)^2
  \check{P}_n(x)\check{P}_m(x)
  =h_n\delta_{nm}\ \ (n,m\in\mathbb{Z}_{\geq 0}),
  \label{orthorel}
\end{equation}
and $\phi_0(x)^2$ is regarded as the weight function.
For the hermiticity of $\cH$, see \cite{os13,os14}.
The systems contain a set of parameters
$\bm{\lambda}=(\lambda_1,\lambda_2,\ldots)$.
The range of $\bm{\lambda}$ is restricted by the hermiticity of $\cH$.
We call the range of parameters for which $\cH$ is hermitian the physical
parameter range of $\bm{\lambda}$ or physical $\bm{\lambda}$.
The parameter dependence is expressed as $f=f(\bm{\lambda})$ and
$f(x)=f(x;\bm{\lambda})$.
The parameter $q$ is $0<q<1$ and $q^{\bm{\lambda}}$ stands for
$q^{(\lambda_1,\lambda_2,\ldots)}=(q^{\lambda_1},q^{\lambda_2},\ldots)$, and
we omit writing $q$-dependence.

The following thirteen polynomials \cite{kls} are studied by the idQM
formulation \cite{os13}:
continuous Hahn (cH),
Meixner-Pollaczek (MP),
Wilson (W),
continuous dual Hahn (cdH),
Askey-Wilson (AW),
continuous dual $q$-Hahn (cd$q$H),
Al-Salam-Chihara (ASC),
continuous big $q$-Hermite (cb$q$He),
continuous $q$-Hermite (c$q$He),
continuous $q$-Jacobi (c$q$J),
continuous $q$-Laguerre (c$q$L),
continuous $q$-Hahn (c$q$H) and
$q$-Meixner-Pollaczek ($q$MP).
These polynomials are classified by the four types of sinusoidal coordinates.
The sinusoidal coordinate $\eta(x)$, the auxiliary function $\varphi(x)$
($\propto\eta(x-i\frac{\gamma}{2})-\eta(x+i\frac{\gamma}{2})$) and
the constant $\gamma$ are \cite{os13}
\begin{equation}
  \begin{array}{rl}
  \text{(\romannumeral1)}:&\eta(x)=x,\quad\varphi(x)=1,\quad\gamma=1,
  \\[5pt]
  \text{(\romannumeral2)}:&\eta(x)=x^2,\quad\varphi(x)=2x,\quad\gamma=1,
  \\[5pt]
  \text{(\romannumeral3)}:&\eta(x)=\cos x,\quad\varphi(x)=2\sin x,\quad
  \gamma=\log q,
  \\[5pt]
  \text{(\romannumeral4)}:&\eta(x;\bm{\lambda})=\cos(x+\phi),\quad
  \varphi(x;\bm{\lambda})=2\sin(x+\phi),\quad\gamma=\log q,
  \end{array}
\end{equation}
and the polynomials are classified as follows:
\begin{equation}
  \begin{array}{rl}
  \text{(\romannumeral1)}:&\text{cH,\,MP},\\
  \text{(\romannumeral2)}:&\text{W,\,cdH},\\
  \text{(\romannumeral3)}:
  &\text{AW,\,cd$q$H,\,ASC,\,cb$q$He,\,c$q$He,\,c$q$J,\,c$q$L},\\
  \text{(\romannumeral4)}:&\text{c$q$H,\,$q$MP}.
  \end{array}
  \label{(i-iv)name}
\end{equation}

These idQM systems satisfy the shape invariance condition \cite{os13,os14,os24}:
\begin{equation}
  \cA(\bm{\lambda})\cA(\bm{\lambda})^{\dagger}
  =\kappa\cA(\bm{\lambda}+\bm{\delta})^{\dagger}
  \cA(\bm{\lambda}+\bm{\delta})+\mathcal{E}_1(\bm{\lambda}),
  \label{shapeinv}
\end{equation}
where $\kappa$ is a real positive constant ($\kappa=1$ for (\romannumeral1),
(\romannumeral2), and $\kappa=q^{-1}$ for (\romannumeral3), (\romannumeral4))
and $\bm{\delta}$ is the shift of the parameters.
This condition combined with the Crum's theorem \cite{os15} allows the
wavefunction $\phi_n(x)$ and energy eigenvalue $\mathcal{E}_n$ of the excited
states to be expressed in terms of the ground state wavefunction $\phi_0(x)$
and the first excited state energy eigenvalue $\mathcal{E}_1$ with shifted
parameters.
As a consequence of the shape invariance, we have
\begin{align}
  \cA(\bm{\lambda})\phi_n(x;\bm{\lambda})
  &=f_n(\bm{\lambda})\phi_{n-1}(x;\bm{\lambda}+\bm{\delta})
  \ \ (n\in\mathbb{Z}_{\geq 0}),
  \label{Aphi=}\\
  \cA(\bm{\lambda})^{\dagger}\phi_{n-1}(x;\bm{\lambda}+\bm{\delta})
  &=b_{n-1}(\bm{\lambda})\phi_n(x;\bm{\lambda})
  \ \ (n\in\mathbb{Z}_{\geq 1}),
  \label{Adphi=}
\end{align}
where $f_n(\bm{\lambda})$ and $b_{n-1}(\bm{\lambda})$ are some constants
satisfying
$f_n(\bm{\lambda})b_{n-1}(\bm{\lambda})=\mathcal{E}_n(\bm{\lambda})$.
These relations can be rewritten as the forward and backward shift relations
for the polynomials:
\begin{align}
  \cF(\bm{\lambda})\check{P}_n(x;\bm{\lambda})
  &=f_n(\bm{\lambda})\check{P}_{n-1}(x;\bm{\lambda}+\bm{\delta})
  \ \ (n\in\mathbb{Z}_{\geq 0}),
  \label{FP=}\\
  \cB(\bm{\lambda})\check{P}_{n-1}(x;\bm{\lambda}+\bm{\delta})
  &=b_{n-1}(\bm{\lambda})\check{P}_n(x;\bm{\lambda})
  \ \ (n\in\mathbb{Z}_{\geq 1}).
  \label{BP=}
\end{align}
Here the forward and backward shift operators $\cF(\bm{\lambda})$ and
$\cB(\bm{\lambda})$ are defined by
\begin{align}
  &\cF(\bm{\lambda})\eqdef
  \phi_0(x;\bm{\lambda}+\bm{\delta})^{-1}\circ
  \cA(\bm{\lambda})\circ\phi_0(x;\bm{\lambda}),
  \label{Fdef}\\
  &\rightarrow\cF(\bm{\lambda})
  =i\varphi(x;\bm{\lambda})^{-1}(e^{\frac{\gamma}{2}p}-e^{-\frac{\gamma}{2}p}),
  \label{F}\\
  &\cB(\bm{\lambda})\eqdef
  \phi_0(x;\bm{\lambda})^{-1}\circ
  \cA(\bm{\lambda})^{\dagger}
  \circ\phi_0(x;\bm{\lambda}+\bm{\delta}),
  \label{Bdef}\\
  &\rightarrow\cB(\bm{\lambda})
  =-i\bigl(V(x;\bm{\lambda})e^{\frac{\gamma}{2}p}
  -V^*(x;\bm{\lambda})e^{-\frac{\gamma}{2}p}\bigr)\varphi(x;\bm{\lambda}),\!
  \label{B}
\end{align}
and satisfy $\wtcH(\bm{\lambda})=\cB(\bm{\lambda})\cF(\bm{\lambda})$.
Let us define the Hilbert space $\mathsf{H}_{\bm{\lambda}}$ as follows:
\begin{equation}
  \mathsf{H}_{\bm{\lambda}}\eqdef
  \text{Span}\bigl[\,\phi_n(x;\bm{\lambda})\bigm|n\in\mathbb{Z}_{\geq 0}\,\bigr].
  \label{Hil}
\end{equation}
The images of the operators $\cA(\bm{\lambda}):
\mathsf{H}_{\bm{\lambda}}\to\mathsf{H}_{\bm{\lambda}+\bm{\delta}}$ and
$\cA(\bm{\lambda})^{\dagger}:\mathsf{H}_{\bm{\lambda}+\bm{\delta}}
\to\mathsf{H}_{\bm{\lambda}}$ are
\begin{equation}
  \text{Im}\,\cA(\bm{\lambda})=\mathsf{H}_{\bm{\lambda}+\bm{\delta}},\quad
  \text{Im}\,\cA(\bm{\lambda})^{\dagger}=
  \text{Span}\bigl[\,\phi_n(x;\bm{\lambda})\bigm|n\in\mathbb{Z}_{\geq 1}\,\bigr].
  \label{ImA}
\end{equation}

\section{Christoffel's Theorem}
\label{sec:CT}

In this section we recall the Christoffel's theorem about a modification
of a measure.

Let $p_n(\eta)$ ($n\in\mathbb{Z}_{\geq 0}$) be the monic orthogonal polynomials
of degree $n$ in $\eta$ with respect to the weight function $w_{\eta}(\eta)$
($\eta_{\text{min}}<\eta<\eta_{\text{max}}$), and $\Phi(\eta)$ be a polynomial
of degree $m$ in $\eta$, which is nonnegative in
$(\eta_{\text{min}},\eta_{\text{max}})$.
We factorize $\Phi(\eta)$ as follows:
\begin{equation}
  \Phi(\eta)=c^{\Phi}\Phi^{\text{monic}}(\eta),\quad
  \Phi^{\text{monic}}(\eta)=\prod_{j=1}^m(\eta-\eta_j),
\end{equation}
where $c^{\Phi}$ is a positive constant.
We assume that $\eta_j$'s are mutually distinct.
Let us define $S_n(\eta)$ ($n\in\mathbb{Z}_{\geq 0}$) as follows:
\begin{align}
  &C_{n,m}\Phi^{\text{monic}}(\eta)S_n(\eta)=
  \left|\begin{array}{cccc}
  p_n(\eta_1)&p_{n+1}(\eta_1)&\cdots&p_{n+m}(\eta_1)\\
  p_n(\eta_2)&p_{n+1}(\eta_2)&\cdots&p_{n+m}(\eta_2)\\
  \vdots&\vdots&\vdots&\vdots\\
  p_n(\eta_m)&p_{n+1}(\eta_m)&\cdots&p_{n+m}(\eta_m)\\
  p_n(\eta)&p_{n+1}(\eta)&\cdots&p_{n+m}(\eta)
  \end{array}\right|,
  \label{Sn}\\
  &C_{n,m}=
  \left|\begin{array}{cccc}
  p_n(\eta_1)&p_{n+1}(\eta_1)&\cdots&p_{n+m-1}(\eta_1)\\
  p_n(\eta_2)&p_{n+1}(\eta_2)&\cdots&p_{n+m-1}(\eta_2)\\
  \vdots&\vdots&\vdots&\vdots\\
  p_n(\eta_m)&p_{n+1}(\eta_m)&\cdots&p_{n+m-1}(\eta_m)\\
  \end{array}\right|.
  \label{Cnm}
\end{align}
Then we have the following theorem.
\begin{thm}\label{thm:Chr}
{\rm [Christoffel] (Theorem 2.7.1 in \cite{ismail})}
$S_n(\eta)$'s are the monic polynomial of degree $n$ in $\eta$ and
they are orthogonal polynomials with respect to the weight function
$\Phi(\eta)w_{\eta}(\eta)$.
\end{thm}
\remark\label{rem:Chr}
If the zero $\eta_j$ has multiplicity $r>1$, the RHS of \eqref{Sn}--\eqref{Cnm}
are replaced as follows:
the corresponding rows of the determinants by derivatives of order
$0,1,\ldots,r-1$ at $\eta_j$.

\remark\label{rem:Uvarov}
A theorem concerning the extension of $\Phi(\eta)$ from a polynomial to
a rational function of $\eta$ was obtained by Uvarov.
See Theorem 2.7.3 in \cite{ismail}.

For the orthogonal polynomials in \S\,\ref{sec:QMF}, we rewrite the formula
\eqref{Sn}.
Let us define $c_n(\bm{\lambda})$ ($n\in\mathbb{Z}_{\geq 0}$),
$x_j$ ($j=1,2,\ldots,m$) and $\check{\Phi}(x;\bm{\lambda})$ as follows:
\begin{align}
  \check{P}_n(x;\bm{\lambda})
  &=c_n(\bm{\lambda})\check{P}^{\text{monic}}_n(x;\bm{\lambda})
  =c_n(\bm{\lambda})
  P^{\text{monic}}_n\bigl(\eta(x;\bm{\lambda});\bm{\lambda}\bigr)
  =P_n\bigl(\eta(x;\bm{\lambda});\bm{\lambda}\bigr),
  \label{cn}\\
  \eta(x_j;\bm{\lambda})&=\eta_j,
  \label{eta(xj)}\\
  \check{\Phi}(x;\bm{\lambda})&\eqdef
  \Phi\bigl(\eta(x;\bm{\lambda});\bm{\lambda}\bigr)
  =c^{\Phi}(\bm{\lambda})\prod_{j=1}^m
  \bigl(\eta(x;\bm{\lambda})-\eta(x_j;\bm{\lambda})\bigr),
  \label{cPhi}
\end{align}
where we omit writing the $\bm{\lambda}$-dependence of $\eta_j$ and $x_j$.
We remark that $x_j$ satisfying \eqref{eta(xj)} may not be unique, but we fix
one value for $x_j$.
Then \eqref{Sn} is rewritten as
\begin{equation}
  \check{\Phi}(x;\bm{\lambda})S_n\bigl(\eta(x;\bm{\lambda});\bm{\lambda}\bigr)
  =\frac{c^{\Phi}(\bm{\lambda})}{c_{n+m}(\bm{\lambda})}
  \frac{\det\bigl(\check{P}_{n+k-1}(x_j;\bm{\lambda})\bigr)_{1\leq j,k\leq m+1}
  \bigm|_{x_{m+1}=x}}
  {\det\bigl(\check{P}_{n+k-1}(x_j;\bm{\lambda})\bigr)_{1\leq j,k\leq m}}
  \ \ (n\in\mathbb{Z}_{\geq 0}).
  \label{Sn2}
\end{equation}
The property of the determinant gives
\begin{equation}
  \det\bigl(\check{P}_{n+\ell_k}(x_j;\bm{\lambda})\bigr)_{1\leq j,k\leq m}
  =\prod_{j=1}^m\check{P}_n(x_j;\bm{\lambda})\times
  D_n^{(\ell_1,\ell_2,\ldots,\ell_m)}(\bm{\lambda}),
\end{equation}
where $n,\ell_k\in\mathbb{Z}_{\geq 0}$ and
$D_n^{(\ell_1,\ell_2,\ldots,\ell_m)}(\bm{\lambda})$ is defined by
\begin{equation}
  D_n^{(\ell_1,\ell_2,\ldots,\ell_m)}(\bm{\lambda})
  \eqdef\det\biggl(\frac{\check{P}_{n+\ell_k}(x_j;\bm{\lambda})}
  {\check{P}_{n}(x_j;\bm{\lambda})}\biggr)_{1\leq j,k\leq m}.
  \label{Dndef}
\end{equation}
The determinant of the denominator of RHS of \eqref{Sn2} is expressed as
\begin{equation}
  \det\bigl(\check{P}_{n+k-1}(x_j;\bm{\lambda})\bigr)_{1\leq j,k\leq m}
  =\prod_{j=1}^m\check{P}_n(x_j;\bm{\lambda})\times
  D_n^{(0,1,\ldots,m-1)}(\bm{\lambda}).
\end{equation}
Expanding the determinant of the numerator of RHS of \eqref{Sn2} in the
$(m+1)$-th row, we obtain the following expression:
\begin{align}
  &\quad\det\bigl(\check{P}_{n+k-1}(x_j;\bm{\lambda})\bigr)_{1\leq j,k\leq m+1}
  \Bigm|_{x_{m+1}=x}\n
  &=\prod_{j=1}^m\check{P}_n(x_j;\bm{\lambda})\times
  \sum_{k=0}^m\check{P}_{n+k}(x;\bm{\lambda})(-1)^{m-k}
  D_n^{(0,1,\ldots,\breve{k},\ldots,m)}(\bm{\lambda}),
\end{align}
where $(0,1,\ldots,\breve{k},\ldots,m)\eqdef(0,1,\ldots,k-1,k+1,\ldots,m)$.
Then \eqref{Sn2} is rewritten as
\begin{equation}
  \check{\Phi}(x;\bm{\lambda})S_n\bigl(\eta(x;\bm{\lambda});\bm{\lambda}\bigr)
  =\frac{c^{\Phi}(\bm{\lambda})}{c_{n+m}(\bm{\lambda})}
  \sum_{k=0}^m\alpha_{n,k}(\bm{\lambda})\check{P}_{n+k}(x;\bm{\lambda})
  \ \ (n\in\mathbb{Z}_{\geq 0}),
  \label{Sn3}
\end{equation}
where $\alpha_{n,k}(\bm{\lambda})$ is given by
\begin{equation}
  \alpha_{n,k}(\bm{\lambda})\eqdef(-1)^{m-k}
  \frac{D_n^{(0,1,\ldots,\breve{k},\ldots,m)}(\bm{\lambda})}
  {D_n^{(0,1,\ldots,m-1)}(\bm{\lambda})}.
  \label{alphank}
\end{equation}

\section{Difference Relations}
\label{sec:DR}

In this section we present some difference relations for the orthogonal
polynomials in \S\,\ref{sec:QMF} by combining the forward shift relations
in \S\,\ref{sec:QMF} and the Christoffel's theorem in \S\,\ref{sec:CT}
with some $\Phi(\eta)$.
In the following we consider the orthogonal polynomials in \eqref{(i-iv)name}
except for c$q$He.

Let us define a function $\check{\Phi}(x;\bm{\lambda})$ as follows:
\begin{equation}
  \check{\Phi}(x;\bm{\lambda})\eqdef
  \kappa^{-1}\varphi(x;\bm{\lambda})^2
  \varphi(x-i\tfrac{\gamma}{2};\bm{\lambda})
  \varphi(x+i\tfrac{\gamma}{2};\bm{\lambda})
  V(x;\bm{\lambda})V^*(x;\bm{\lambda}).
  \label{cPhidef}
\end{equation}
By using explicit forms of $V(x;\bm{\lambda})$, $\varphi(x;\bm{\lambda})$,
$\gamma$ and $\kappa$ (see \S\,\ref{sec:ExAW} and Appendix \ref{app:ExidQM}),
a direct calculation shows the following proposition.
\begin{prop}\label{prop:Phi}
The function $\check{\Phi}(x;\bm{\lambda})$ \eqref{cPhidef} is a polynomial of
degree $m$ in $\eta=\eta(x;\bm{\lambda})$,
\eqref{cPhi},
\begin{equation}
  m=\left\{
  \begin{array}{ll}
  4&:\text{\rm cH,\,W,\,AW,\,c$q$J,\,c$q$H}\\
  3&:\text{\rm cdH,\,cd$q$H}\\
  2&:\text{\rm MP,\,ASC,\,c$q$L,\,$q$MP}\\
  1&:\text{\rm cb$q$He}
  \end{array}\right..
\end{equation}
\end{prop}
\remark\label{rem:cqHe}
For c$q$He, $\check{\Phi}(x;\bm{\lambda})$ \eqref{cPhidef} becomes a constant,
$\check{\Phi}(x;\bm{\lambda})=1$.

This $\check{\Phi}(x)$ relates $\phi_0(x)$ with different
parameters in the following way.
\begin{prop}\label{prop:Phiphi02}
$\check{\Phi}(x)$ and $\phi_0(x)$ satisfy the following relation:
\begin{equation}
  \check{\Phi}(x;\bm{\lambda})\phi_0(x;\bm{\lambda})^2
  =\phi_0(x;\bm{\lambda}+2\bm{\delta})^2.
  \label{Phiphi02=}
\end{equation}
\end{prop}
Proof: By using explicit forms of $\check{\Phi}(x;\bm{\lambda})$ and
$\phi_0(x;\bm{\lambda})$, a direct calculation shows \eqref{Phiphi02=}.
\hfill\fbox{}

The square of the ground state wavefunctions, $\phi_0(x;\bm{\lambda})^2$ and
$\phi_0(x;\bm{\lambda}+2\bm{\delta})^2$, are the weight functions of the
orthogonal polynomials, $\check{P}_n(x;\bm{\lambda})$ and
$\check{P}_n(x;\bm{\lambda}+2\bm{\delta})$, respectively.
Therefore the orthogonal polynomial
$S_n\bigl(\eta(x;\bm{\lambda});\bm{\lambda}\bigr)$ in Theorem\,\ref{thm:Chr},
\eqref{Sn3}, is given by
\begin{equation}
  S_n\bigl(\eta(x;\bm{\lambda});\bm{\lambda}\bigr)
  =\check{P}^{\text{monic}}_n(x;\bm{\lambda}+2\bm{\delta})
  =c_n(\bm{\lambda}+2\bm{\delta})^{-1}\check{P}_n(x;\bm{\lambda}+2\bm{\delta}).
\end{equation}
Then Theorem\,\ref{thm:Chr}, \eqref{Sn3}, gives the following theorem.
\begin{thm}\label{thm:PhiPn}
We have the following relations,
\begin{equation}
  \check{\Phi}(x;\bm{\lambda})\check{P}_n(x;\bm{\lambda}+2\bm{\delta})
  =\beta_n(\bm{\lambda})\sum_{k=0}^m\alpha_{n,k}(\bm{\lambda})
  \check{P}_{n+k}(x;\bm{\lambda})
  \ \ (n\in\mathbb{Z}_{\geq 0}),
  \label{PhiPn=}
\end{equation}
where $\alpha_{n,k}(\bm{\lambda})$ is given by \eqref{alphank} and
$\beta_n(\bm{\lambda})$ is given by
\begin{equation}
  \beta_n(\bm{\lambda})\eqdefrm c^{\Phi}(\bm{\lambda})
  \frac{c_n(\bm{\lambda}+2\bm{\delta})}{c_{n+m}(\bm{\lambda})}.
  \label{betan}
\end{equation}
Explicit expressions of $\check{\Phi}(x;\bm{\lambda})$,
$\alpha_{n,k}(\bm{\lambda})$ and $\beta_n(\bm{\lambda})$ are given in
\S\,\ref{sec:ExAW} and Appendix \ref{app:ExidQM}.
\end{thm}
\remark\label{rem:MP}
For MP, this formula is obtained explicitly in \cite{is25} (eq.(21) in
\cite{is25}). For its explicit form, see \S\,\ref{app:MP}.

This theorem relates $\check{P}_n(x)$'s with the parameters $\bm{\lambda}$ and
$\bm{\lambda}+2\bm{\delta}$.
By using the forward shift relation, let us obtain some relation among
$\check{P}_n(x)$'s with $\bm{\lambda}$ only.
{}From the forward shift relation \eqref{FP=}, we have
\begin{equation}
  \cF(\bm{\lambda}+\bm{\delta})\cF(\bm{\lambda})
  \check{P}_{n+2}(x;\bm{\lambda})
  =f_{n+1}(\bm{\lambda}+\bm{\delta})f_{n+2}(\bm{\lambda})
  \check{P}_n(x;\bm{\lambda}+2\bm{\delta}).
  \label{FFP}
\end{equation}
Explicit form of the forward shift operator \eqref{F} gives the following
proposition.
\begin{prop}\label{prop:FF}
The product of $\cF(\bm{\lambda}+\bm{\delta})$ and $\cF(\bm{\lambda})$ is
expressed as
\begin{align}
  &\quad\cF(\bm{\lambda}+\bm{\delta})\cF(\bm{\lambda})\n
  &=\frac{1}{\varphi(x-i\frac{\gamma}{2};\bm{\lambda})
  \varphi(x+i\frac{\gamma}{2};\bm{\lambda})}\Bigl(
  \kappa^{\frac12}+\kappa^{-\frac12}
  -\frac{\varphi(x+i\frac{\gamma}{2};\bm{\lambda})}{\varphi(x;\bm{\lambda})}
  e^{\gamma p}
  -\frac{\varphi(x-i\frac{\gamma}{2};\bm{\lambda})}{\varphi(x;\bm{\lambda})}
  e^{-\gamma p}\Bigr).
  \label{FF=}
\end{align}
\end{prop}
Proof: First we remark
$\varphi(x;\bm{\lambda}+\bm{\delta})=\varphi(x;\bm{\lambda})$.
By using explicit forms of $\varphi(x;\bm{\lambda})$ and $\gamma$, we have
\begin{equation}
  \varphi(x-i\tfrac{\gamma}{2};\bm{\lambda})
  +\varphi(x+i\tfrac{\gamma}{2};\bm{\lambda})
  =(\kappa^{\frac12}+\kappa^{-\frac12})\varphi(x;\bm{\lambda}).
\end{equation}
A direct calculation using this relation and \eqref{F} shows \eqref{FF=}.
\hfill\fbox{}

Combining \eqref{FFP}, Proposition\,\ref{prop:FF},
Theorem\,\ref{thm:PhiPn} and \eqref{cPhidef}, we obtain the following theorem.
\begin{thm}\label{thm:sabunrel}
We have the following difference relations for $\check{P}_n(x;\bm{\lambda})$'s, 
\begin{align}
  &\quad\kappa^{-1}\varphi(x;\bm{\lambda})V(x;\bm{\lambda})V^*(x;\bm{\lambda})
  \bigl((\kappa^{\frac12}+\kappa^{-\frac12})\varphi(x;\bm{\lambda})
  \check{P}_{n+2}(x;\bm{\lambda})\n
  &\qquad\qquad\qquad
  -\varphi(x+i\tfrac{\gamma}{2};\bm{\lambda})
  \check{P}_{n+2}(x-i\gamma;\bm{\lambda})
  -\varphi(x-i\tfrac{\gamma}{2};\bm{\lambda})
  \check{P}_{n+2}(x+i\gamma;\bm{\lambda})\bigr)
  \label{sabunrel}\\
  &=\beta^{\cF}_n(\bm{\lambda})\sum_{k=0}^m\alpha_{n,k}(\bm{\lambda})
  \check{P}_{n+k}(x;\bm{\lambda})
  \ \ (n\in\mathbb{Z}_{\geq 0}),\nonumber
\end{align}
where $\beta^{\cF}_n(\bm{\lambda})$ is given by
\begin{equation}
  \beta^{\cF}_n(\bm{\lambda})\eqdefrm
  f_{n+1}(\bm{\lambda}+\bm{\delta})f_{n+2}(\bm{\lambda})\beta_n(\bm{\lambda}).
  \label{betaFn}
\end{equation}
\end{thm}

{}From Proposition\,\ref{prop:Phiphi02}, we have the following property.
\begin{thm}\label{thm:map}
Multiplying by $\sqrt{\check{\Phi}(x;\bm{\lambda})}$ is a surjective map
from $\mathsf{H}_{\bm{\lambda}+2\bm{\delta}}$ to $\mathsf{H}_{\bm{\lambda}}$,
\begin{equation}
  \sqrt{\check{\Phi}(x;\bm{\lambda})}:
  \mathsf{H}_{\bm{\lambda}+2\bm{\delta}}\to\mathsf{H}_{\bm{\lambda}},\quad
  \text{\rm Im}\,\sqrt{\check{\Phi}(x;\bm{\lambda})}
  =\mathsf{H}_{\bm{\lambda}}.
  \label{srPhi}
\end{equation} 
\end{thm}
Proof:
By using \eqref{phin=}, \eqref{Phiphi02=} and \eqref{PhiPn=},
we obtain
\begin{align}
  \sqrt{\check{\Phi}(x;\bm{\lambda})}\,\phi_n(x;\bm{\lambda}+2\bm{\delta})
  &=\check{\Phi}(x;\bm{\lambda})
  \frac{\phi_0(x;\bm{\lambda})}{\phi_0(x;\bm{\lambda}+2\bm{\delta})}\cdot
  \phi_0(x;\bm{\lambda}+2\bm{\delta})\check{P}_n(x;\bm{\lambda}+2\bm{\delta})\n
  &=\phi_0(x;\bm{\lambda})\check{\Phi}_n(x;\bm{\lambda})
  \check{P}_n(x;\bm{\lambda}+2\bm{\delta})\n
  &=\phi_0(x;\bm{\lambda})\beta_n(\bm{\lambda})
  \sum_{k=0}^m\alpha_{n,k}(\bm{\lambda})\check{P}_{n+k}(x;\bm{\lambda})\\
  &=\beta_n(\bm{\lambda})
  \sum_{k=0}^m\alpha_{n,k}(\bm{\lambda})\phi_{n+k}(x;\bm{\lambda})
  \in\mathsf{H}_{\bm{\lambda}}.
  \nonumber
\end{align}
By setting $\psi_n(x;\bm{\lambda})=\sqrt{\check{\Phi}(x;\bm{\lambda})}\,
\phi_n(x;\bm{\lambda}+2\bm{\delta})$, this relation is expressed as a matrix
form,
\begin{equation}
  \begin{pmatrix}
  \psi_0(x;\bm{\lambda})\\
  \psi_1(x;\bm{\lambda})\\
  \vdots
  \end{pmatrix}
  =
  \begin{pmatrix}
  \hspace{-9mm}\beta_0\alpha_{0,0}\ \cdots\ \beta_0\alpha_{0,m}\qquad\\
  \hspace{12mm}\beta_1\alpha_{1,0}\ \cdots\ \beta_1\alpha_{1,m}\qquad\\
  \hspace{19mm}\ddots\ \ \ \cdots\ \ \ \ddots\qquad
  \end{pmatrix}
  \begin{pmatrix}
  \phi_0(x;\bm{\lambda})\\
  \phi_1(x;\bm{\lambda})\\
  \vdots
  \end{pmatrix},
\end{equation}
where we omit writing the $\bm{\lambda}$-dependence of $\alpha_{n,k}$ and
$\beta_n$.
The diagonal elements of this matrix are $\beta_n\alpha_{n,0}$, and we can
check that they do not vanish for the physical $\bm{\lambda}$.
So this matrix has an inverse.
Thus $\phi_n(x;\bm{\lambda})$ can be expressed as a linear combination of
$\psi_n(x;\bm{\lambda})$.
\hfill\fbox{}

\remark\label{rem:trivial}
Since the elements of $\mathsf{H}_{\bm{\lambda}}$ are expressed as
$\sum\limits_{n=0}^{\infty}c_n\phi_n(x;\bm{\lambda})=\phi_0(x;\bm{\lambda})
\sum\limits_{n=0}^{\infty}c_n\check{P}_n(x;\bm{\lambda})$,
$\mathsf{H}_{\bm{\lambda}}$ is essentially characterized by
$\phi_0(x;\bm{\lambda})$.
Any polynomial can be expanded by other polynomials, e.g.
$\check{P}_n(x;\bm{\lambda}+2\bm{\delta})
=\sum\limits_{k=0}^nc_{n,k}\check{P}_k(x;\bm{\lambda})$.
So multiplying by
\begin{equation}
  \frac{1}{\sqrt{\check{\Phi}(x;\bm{\lambda})}}
  =\frac{\phi_0(x;\bm{\lambda})}{\phi_0(x;\bm{\lambda}+2\bm{\delta})}
\end{equation}
is also a surjective map from $\mathsf{H}_{\bm{\lambda}+2\bm{\delta}}$ to
$\mathsf{H}_{\bm{\lambda}}$,
\begin{align}
  \frac{1}{\sqrt{\check{\Phi}(x;\bm{\lambda})}}
  \phi_n(x;\bm{\lambda}+2\bm{\delta})
  &=\phi_0(x;\bm{\lambda})\check{P}_n(x;\bm{\lambda}+2\bm{\delta})
  \nonumber\\[-12pt]
  &=\phi_0(x;\bm{\lambda})\sum_{k=0}^nc_{n,k}\check{P}_k(x;\bm{\lambda})
  =\sum_{k=0}^nc_{n,k}\phi_k(x;\bm{\lambda})
  \in\mathsf{H}_{\bm{\lambda}}.
\end{align}
In the same sense,
$\check{\Phi}(x;\bm{\lambda})\check{P}_n(x;\bm{\lambda}+2\bm{\delta})$ is
expanded as $\sum\limits_{k=0}^{n+m}c'_{n,k}\check{P}_k(x;\bm{\lambda})$.
Therefore, Theorem\,\ref{thm:map} is shown by Proposition\,\ref{prop:Phi} and
Proposition\,\ref{prop:Phiphi02} (namely,
$\frac{\phi_0(x;\bm{\lambda}+2\bm{\delta})^2}{\phi_0(x;\bm{\lambda})^2}$
is a polynomial in $\eta(x;\bm{\lambda})$) without using
Theorem\,\ref{thm:PhiPn}.

\remark\label{rem:range}
This Theorem\,\ref{thm:map} holds only for physical $\bm{\lambda}$.
On the other hand,
\eqref{FP=}--\eqref{BP=}, Theorem\,\ref{thm:PhiPn} and
Theorem\,\ref{thm:sabunrel}
hold for any $\bm{\lambda}$, because they are algebraic relations.

\subsection{Explicit expression : AW case}
\label{sec:ExAW}

As an illustration, we present explicit expressions of various quantities
for AW case. For other cases, see Appendix\,\ref{app:ExidQM}.

A set of parameters $\bm{\lambda}$ is
\begin{equation}
  q^{\bm{\lambda}}=(a_1,a_2,a_3,a_4)\ \ (a_j\in\mathbb{C}),\quad
  \{a_1^*,a_2^*,a_3^*,a_4^*\}=\{a_1,a_2,a_3,a_4\}\ (\text{as a set}),
\end{equation}
and $b_j$ ($j=1,2,3,4$) are defined by (please do not confuse these $b_j$ and
$b_n(\bm{\lambda})$ in \eqref{BP=})
\begin{equation}
  b_1\eqdef\sum_{j=1}^4a_j,\quad
  b_2\eqdef\!\!\sum_{1\leq j<k\leq 4}\!\!a_ja_k,\quad
  b_3\eqdef\!\!\!\!\sum_{1\leq j<k<l\leq 4}\!\!\!\!a_ja_ka_l,\quad
  b_4\eqdef a_1a_2a_3a_4.
 \label{bj}
%
%
\end{equation}
The basic data of the idQM system of AW type are as follows \cite{os13}:
\begin{align}
  &x_{\text{min}}=0,\quad x_{\text{max}}=\pi,\quad\gamma=\log q,\\
  &\eta(x)=\cos x,\quad\varphi(x)=2\sin x,\quad\kappa=q^{-1},\quad
  \bm{\delta}=(\tfrac12,\tfrac12,\tfrac12,\tfrac12),\\
  &V(x;\bm{\lambda})=\frac{(1-a_1e^{ix})(1-a_2e^{ix})(1-a_3e^{ix})(1-a_4e^{ix})}
  {(1-e^{2ix})(1-qe^{2ix})},\\
  &\mathcal{E}_n(\bm{\lambda})=(q^{-n}-1)(1-b_4q^{n-1}),
  \ \ f_n(\bm{\lambda})=q^{\frac{n}{2}}(q^{-n}-1)(1-b_4q^{n-1}),
  \ \ b_n(\bm{\lambda})=q^{-\frac{n+1}{2}},\\
  &\phi_0(x;\bm{\lambda})^2
  =\frac{(e^{2ix},e^{-2ix};q)_{\infty}}
  {\prod_{j=1}^4(a_je^{ix},a_je^{-ix};q)_{\infty}},\quad
  h_n(\bm{\lambda})
  =\frac{2\pi(b_4q^{n-1};q)_n(b_4q^{2n};q)_{\infty}}
  {(q^{n+1};q)_{\infty}\prod_{1\leq i<j\leq 4}(a_ia_jq^n;q)_{\infty}},\\
  &\check{P}_n(x;\bm{\lambda})
  =P_n\bigl(\eta(x);\bm{\lambda}\bigr)
  =p_n\bigl(\eta(x);a_1,a_2,a_3,a_4|q\bigr)\n
  &\phantom{\check{P}_n(x;\bm{\lambda})}
  =\frac{(a_1a_2,a_1a_3,a_1a_4\,;q)_n}{a_1^n}
  {}_4\phi_3\Bigl(
  \genfrac{}{}{0pt}{}{q^{-n},\,b_4q^{n-1},\,a_1e^{ix},\,a_1e^{-ix}}
  {a_1a_2,\,a_1a_3,\,a_1a_4}\Bigm|q,q\Bigr),
  \label{cPn:AW}\\
  &c_n(\bm{\lambda})=2^n(b_4q^{n-1};q)_n,
\end{align}
where $p_n(\eta;a_1,a_2,a_3,a_4|q)$ is the Askey-Wilson polynomial \cite{kls}.
We remark that $\check{P}_n(x;\bm{\lambda})$ is invariant under the
permutations of $a_j$'s, and
$\check{P}_n(-x;\bm{\lambda})=\check{P}_n(x;\bm{\lambda})$.
The physical $\bm{\lambda}$ is $|a_j|<1$ ($j=1,2,3,4$).

{}From \eqref{cPhidef}, we have
\begin{equation}
  \check{\Phi}(x;\bm{\lambda})=\prod_{j=1}^4(1-a_je^{ix})(1-a_je^{-ix}),\quad
  m=4,\quad c^{\Phi}(\bm{\lambda})=16b_4,
\end{equation}
which is a polynomial in $\eta(x)=\cos x$ by
$(1-ae^{ix})(1-ae^{-ix})=1+a^2-2a\cos x$.
The zeros $\eta_j$ and $x_j$ ($j=1,2,3,4$) are
\begin{equation}
  \eta_j=\frac{a_j+a_j^{-1}}{2},\quad e^{ix_j}=a_j,
\end{equation}
where we choose one value for $x_j$ as mentioned below \eqref{cPhi}
($e^{ix_j}=a_j^{-1}$ also gives the same $\eta_j$).
The expression \eqref{cPn:AW} gives
$\check{P}_n(x_1;\bm{\lambda})=a_1^{-n}(a_1a_2,a_1a_3,a_1a_4;q)_n$ and
the symmetry under the permutations of $a_j$'s gives
\begin{equation}
  \check{P}_n(x_j;\bm{\lambda})
  =a_j^{-n}\prod_{\substack{k=1\\k\neq j}}^4(a_ja_k;q)_n
  =\frac{\prod_{k=1}^4(a_ja_k;q)_n}{a_j^n(a_j^2;q)_n}
  \ \ (j=1,2,3,4).
  \label{cPnxj}
\end{equation}
By direct calculation, we can obtain
$D_n^{(0,1,\ldots,\breve{k},\ldots,4)}(\bm{\lambda})$ ($k=0,1,2,3,4$),
\eqref{Dndef}.
Here we write down $D_n^{(0,1,2,3)}(\bm{\lambda})$ and
$\alpha_{n,k}(\bm{\lambda})$ ($k=0,1,2,3,4$) \eqref{alphank}
($D_n^{(0,1,\ldots,\breve{k},\ldots,4)}(\bm{\lambda})$ ($k=0,1,2,3$) are
obtained as
$(-1)^{4-k}\alpha_{n,k}(\bm{\lambda})D_n^{(0,1,\ldots,3)}(\bm{\lambda})$),
\begin{align}
  &D_n^{(0,1,2,3)}(\bm{\lambda})
  =\prod_{1\leq j<k\leq 4}(a_j-a_k)(1-a_ja_kq^n)\cdot
  b_4^{-3}(b_4q^{2n},b_4q^{2n+2};q)_3,\\
  &\alpha_{n,4}(\bm{\lambda})=1,
  \label{alphan4:AW}\\
  &\alpha_{n,3}(\bm{\lambda})=-\frac{1-b_4q^{2n+5}}{1-b_4q^{2n+2}}\,b_4^{-1}
  \bigl(b_3-(1+q+q^2)(b_1-b_3q^{n+2})b_4q^{n+1}-b_1b_4^2q^{3n+6}\bigr),\\
  &\alpha_{n,2}(\bm{\lambda})=
  \frac{(1-b_4q^{2n+3})(1-b_4q^{2n+6})}{(1-b_4q^{2n+1})(1-b_4q^{2n+2})}
  \,b_4^{-1}\Bigl(b_2(1-b_4q^{2n+2})(1+b_4q^{2n+3})(1-b_4q^{2n+4})\n
  &\phantom{\alpha_{n,2}(\bm{\lambda})=}\quad
  +(1+q+q^2)(b_3^2+b_1^2b_4)(1+b_4q^{2n+3})q^{2n+2}\n
  &\phantom{\alpha_{n,2}(\bm{\lambda})=}\quad
  -(1+q)\bigl(qb_1b_3+(1+q^2)b_4\bigl)(1+b_4^2q^{4n+6})q^n\n
  &\phantom{\alpha_{n,2}(\bm{\lambda})=}\quad
  -(1+q)\bigl(q(1+q)^2b_1b_3-(1+q^2)^2b_4\bigr)b_4q^{3n+2}\Bigr),\\
  &\alpha_{n,1}(\bm{\lambda})=-\prod_{1\leq j<k\leq 4}(1-a_ja_kq^{n+1})\cdot
  \frac{(1-b_4q^{2n+5})(1-b_4q^{2n+6})}{(1-b_4q^{2n})(1-b_4q^{2n+2})}\n
  &\phantom{\alpha_{n,1}(\bm{\lambda})=}\quad\times
  b_4^{-1}\bigl(b_1-(1+q+q^2)(b_3-b_1b_4q^{n+1})q^n-b_3b_4q^{3n+3}\bigr),\\
  &\alpha_{n,0}(\bm{\lambda})=\prod_{1\leq j<k\leq 4}(a_ja_kq^n;q)_2\cdot
  \frac{(b_4q^{2n+4};q)_3}{(b_4q^{2n};q)_3}\,b_4^{-1}.
  \label{alphan0:AW}
\end{align}
The constant $\beta_n(\bm{\lambda})$ \eqref{betan} and
$\beta^{\cF}_n(\bm{\lambda})$ \eqref{betaFn} are
\begin{equation}
  \beta_n(\bm{\lambda})=\frac{b_4}{(b_4q^{2n+3};q)_4},\quad
  \beta^{\cF}_n(\bm{\lambda})
  =\frac{b_4q^{-n-\frac32}(q^{n+1},b_4q^{n+1};q)_2}{(b_4q^{2n+3};q)_4}.
  \label{betan:AW}
\end{equation}

\subsubsection{shift with only one parameter}
\label{sec:shiftone}

In order to obtain Theorem\,\ref{thm:sabunrel} for AW case, we have shifted
all $a_j$'s by $1$ in Theorem\,\ref{thm:PhiPn}.
But we can consider shifting only one $a_j$ by $1$ and obtain theorems similar
to Theorem\,\ref{thm:PhiPn} and Theorem\,\ref{thm:map}.
Such one $a_j$ shift was considered for Wilson polynomial in \cite{jsj14}.

For each $j$ ($j=1,2,3,4$), let us define $\bm{\delta}^{(j)}$ and
$\check{\Phi}^{(j)}(x;\bm{\lambda})$ as follows:
\begin{align}
  &\bm{\delta}^{(1)}\eqdef(\tfrac12,0,0,0),
  \ \ \bm{\delta}^{(2)}\eqdef(0,\tfrac12,0,0),
  \ \ \bm{\delta}^{(3)}\eqdef(0,0,\tfrac12,0),
  \ \ \bm{\delta}^{(4)}\eqdef(0,0,0,\tfrac12),\\
  &\check{\Phi}^{(j)}(x;\bm{\lambda})\eqdef(1-a_je^{ix})(1-a_je^{-ix}),
\end{align}
where we assume $a_j^*=a_j$ for the reality of
$\check{\Phi}^{(j)}(x;\bm{\lambda})$
($\check{\Phi}^{(j)}(x;\bm{\lambda})\in\mathbb{R}$ for $x\in\mathbb{R}$).
Then we have
\begin{equation}
  \check{\Phi}^{(j)}(x;\bm{\lambda})
  =\Phi^{(j)}\bigl(\eta(x);\bm{\lambda}\bigr),\quad
  m=1,\quad c^{\Phi^{(j)}}(\bm{\lambda})=-2a_j,
\end{equation}
and
\begin{equation}
  \check{\Phi}^{(j)}(x;\bm{\lambda})\phi_0(x;\bm{\lambda})^2
  =\phi_0(x;\bm{\lambda}+2\bm{\delta}^{(j)})^2.
\end{equation}
This relation implies that the orthogonal polynomials
$S_n(\eta(x);\bm{\lambda})$ in Theorem\,\ref{thm:Chr}, \eqref{Sn3}, are given by
\begin{equation}
  S_n(\eta(x);\bm{\lambda})
  =\check{P}^{\text{monic}}_n(x;\bm{\lambda}+2\bm{\delta}^{(j)})
  =c_n(\bm{\lambda}+2\bm{\delta}^{(j)})^{-1}
  \check{P}_n(x;\bm{\lambda}+2\bm{\delta}^{(j)}).
\end{equation}
The zeros of $\check{\Phi}^{(j)}(x;\bm{\lambda})$, $\eta^{(j)}_1$ and
$x^{(j)}_1$, and $\check{P}_n(x^{(j)};\bm{\lambda})$ are
\begin{equation}
  \eta^{(j)}_1=\eta_j,\quad x^{(j)}_1=x_j,\quad
  \check{P}_n(x^{(j)}_1;\bm{\lambda})=\check{P}_n(x_j;\bm{\lambda}).
\end{equation}
Then Theorem\,\ref{thm:Chr}, \eqref{Sn3}, gives the following relations:
\begin{equation}
  \check{\Phi}^{(j)}(x;\bm{\lambda})
  \check{P}_n(x;\bm{\lambda}+2\bm{\delta}^{(j)})
  =\beta^{(j)}_n(\bm{\lambda})\sum_{k=0}^1\alpha^{(j)}_{n,k}(\bm{\lambda})
  \check{P}_{n+k}(x;\bm{\lambda})
  \ \ (n\in\mathbb{Z}_{\geq 0}),
  \label{PhijPn=}
\end{equation}
where $\alpha^{(j)}_{n,k}(\bm{\lambda})$ and $\beta^{(j)}_n(\bm{\lambda})$ are
 given by
\begin{align}
  \alpha^{(j)}_{n,1}(\bm{\lambda})&=1,\\
  \alpha^{(j)}_{n,0}(\bm{\lambda})
  &=-\frac{D_n^{(1)}(\bm{\lambda})}{D_n^{(0)}(\bm{\lambda})}
  =-D_n^{(1)}(\bm{\lambda})
  =-\frac{\check{P}_{n+1}(x^{(j)}_1;\bm{\lambda})}
  {\check{P}_n(x^{(j)}_1;\bm{\lambda})}
  =-\frac{\prod_{k=1}^4(1-a_ja_kq^n)}{a_j(1-a_j^2q^n)},
  \label{alphajnk}\\
  \beta^{(j)}_n(\bm{\lambda})&=c^{\Phi^{(j)}}(\bm{\lambda})
  \frac{c_n(\bm{\lambda}+2\bm{\delta}^{(j)})}{c_{n+1}(\bm{\lambda})}
  =\frac{-a_j}{1-b_4q^{2n}}.
  \label{betajn}
\end{align}
Note that the relations \eqref{PhijPn=} hold for $a_j\in\mathbb{C}$.
As in Theorem\,\ref{thm:map}, multiplying by
$\sqrt{\check{\Phi}^{(j)}(x;\bm{\lambda})}$ is a surjective map
from $\mathsf{H}_{\bm{\lambda}+2\bm{\delta}^{(j)}}$ to
$\mathsf{H}_{\bm{\lambda}}$ for physical $\bm{\lambda}$,
\begin{equation}
  \sqrt{\check{\Phi}^{(j)}(x;\bm{\lambda})}:
  \mathsf{H}_{\bm{\lambda}+2\bm{\delta}^{(j)}}\to\mathsf{H}_{\bm{\lambda}},\quad
  \text{\rm Im}\,\sqrt{\check{\Phi}^{(j)}(x;\bm{\lambda})}
  =\mathsf{H}_{\bm{\lambda}}.
  \label{srPhij}
\end{equation}

We remark that $\check{\Phi}(x;\bm{\lambda})$ is expressed as
\begin{equation}
  \check{\Phi}(x;\bm{\lambda})
  =\prod_{j=1}^4\check{\Phi}^{(j)}(x;\bm{\lambda}),\quad
  c^{\Phi}(\bm{\lambda})=\prod_{j=1}^4c^{\Phi^{(j)}}(\bm{\lambda}),
\end{equation}
and $\check{\Phi}^{(j)}(x;\bm{\lambda})$ has the following property,
\begin{equation}
  \check{\Phi}^{(j)}(x;\bm{\lambda}+\bm{\delta}^{(k)})
  =\check{\Phi}^{(j)}(x;\bm{\lambda})\ \ (k\neq j),
\end{equation}
because $\check{\Phi}^{(j)}(x;\bm{\lambda})$ does not depend on $a_k$
($k\neq j$).
So, Theorem\,\ref{thm:PhiPn} and Theorem\,\ref{thm:map} can be obtained by
repeated use of the relations \eqref{PhijPn=} and \eqref{srPhij} respectively.
For example, by expressing $\check{\Phi}(x;\bm{\lambda})$ as
\begin{align}
  &\check{\Phi}(x;\bm{\lambda})=\check{\Phi}^{(1)}(x;\bm{\lambda})
  \check{\Phi}^{(2)}(x;\bm{\lambda}+2\bm{\delta}^{(1)})
  \check{\Phi}^{(3)}(x;\bm{\lambda}+2\bm{\delta}^{(12)})
  \check{\Phi}^{(4)}(x;\bm{\lambda}+2\bm{\delta}^{(123)}),\n
  &\quad\bigl(\bm{\delta}^{(12)}\eqdef\bm{\delta}^{(1)}+\bm{\delta}^{(2)},
  \ \ \bm{\delta}^{(123)}\eqdef\bm{\delta}^{(1)}+\bm{\delta}^{(2)}
  +\bm{\delta}^{(3)}\bigr),
\end{align}
$\alpha_{n,k}(\bm{\lambda})$ and $\beta_n(\bm{\lambda})$ are expressed in terms
of $\alpha^{(j)}_{n,k}(\bm{\lambda})$ and $\beta^{(j)}_n(\bm{\lambda})$ as
follows:
\begin{align}
  \alpha_{n,4}(\bm{\lambda})&=1,
  \label{alphan4:AW2}\\
  \alpha_{n,3}(\bm{\lambda})&=
  \tilde{\alpha}^{(1)}_{n+3}(\bm{\lambda})
  +\tilde{\alpha}^{(2)}_{n+2}(\bm{\lambda})
  +\tilde{\alpha}^{(3)}_{n+1}(\bm{\lambda})
  +\tilde{\alpha}^{(4)}_{n}(\bm{\lambda}),\\
  \alpha_{n,2}(\bm{\lambda})&=
  \tilde{\alpha}^{(1)}_{n+2}(\bm{\lambda})
  \tilde{\alpha}^{(2)}_{n+2}(\bm{\lambda})
  +\tilde{\alpha}^{(1)}_{n+2}(\bm{\lambda})
  \tilde{\alpha}^{(3)}_{n+1}(\bm{\lambda})
  +\tilde{\alpha}^{(1)}_{n+2}(\bm{\lambda})
  \tilde{\alpha}^{(4)}_n(\bm{\lambda})\n
  &\quad
  +\tilde{\alpha}^{(2)}_{n+1}(\bm{\lambda})
  \tilde{\alpha}^{(3)}_{n+1}(\bm{\lambda})
  +\tilde{\alpha}^{(2)}_{n+1}(\bm{\lambda})
  \tilde{\alpha}^{(4)}_n(\bm{\lambda})
  +\tilde{\alpha}^{(3)}_n(\bm{\lambda})
  \tilde{\alpha}^{(4)}_n(\bm{\lambda}),\\
  \alpha_{n,1}(\bm{\lambda})&=
  \tilde{\alpha}^{(1)}_{n+1}(\bm{\lambda})
  \tilde{\alpha}^{(2)}_{n+1}(\bm{\lambda})
  \tilde{\alpha}^{(3)}_{n+1}(\bm{\lambda})
  +\tilde{\alpha}^{(1)}_{n+1}(\bm{\lambda})
  \tilde{\alpha}^{(2)}_{n+1}(\bm{\lambda})
  \tilde{\alpha}^{(4)}_n(\bm{\lambda})\n
  &\quad
  +\tilde{\alpha}^{(1)}_{n+1}(\bm{\lambda})
  \tilde{\alpha}^{(3)}_n(\bm{\lambda})
  \tilde{\alpha}^{(4)}_n(\bm{\lambda})
  +\tilde{\alpha}^{(2)}_n(\bm{\lambda})
  \tilde{\alpha}^{(3)}_n(\bm{\lambda})
  \tilde{\alpha}^{(4)}_n(\bm{\lambda}),\\
  \alpha_{n,0}(\bm{\lambda})&=
  \tilde{\alpha}^{(1)}_n(\bm{\lambda})\tilde{\alpha}^{(2)}_n(\bm{\lambda})
  \tilde{\alpha}^{(3)}_n(\bm{\lambda})\tilde{\alpha}^{(4)}_n(\bm{\lambda}),
  \label{alphan0:AW2}\\
  \beta_n(\bm{\lambda})
  &=\beta^{(1)}_{n+3}(\bm{\lambda})
  \beta^{(2)}_{n+2}(\bm{\lambda}+2\bm{\delta}^{(1)})
  \beta^{(3)}_{n+1}(\bm{\lambda}+2\bm{\delta}^{(12)})
  \beta^{(4)}_n(\bm{\lambda}+2\bm{\delta}^{(123)}),
  \label{betan:AW2}
\end{align}
where $\tilde{\alpha}^{(j)}_n(\bm{\lambda})$ are given by
\begin{align}
  \tilde{\alpha}^{(1)}_n(\bm{\lambda})
  &\eqdef\alpha^{(1)}_{n,0}(\bm{\lambda}),
  \label{talpha1n}\\
  \tilde{\alpha}^{(2)}_n(\bm{\lambda})
  &\eqdef\alpha^{(2)}_{n,0}(\bm{\lambda}+2\bm{\delta}^{(1)})
  \frac{\beta^{(1)}_n(\bm{\lambda})}{\beta^{(1)}_{n+1}(\bm{\lambda})},\\
  \tilde{\alpha}^{(3)}_n(\bm{\lambda})
  &\eqdef\alpha^{(3)}_{n,0}(\bm{\lambda}+2\bm{\delta}^{(12)})
  \frac{\beta^{(1)}_{n+1}(\bm{\lambda})}{\beta^{(1)}_{n+2}(\bm{\lambda})}
  \frac{\beta^{(2)}_n(\bm{\lambda}+2\bm{\delta}^{(1)})}
  {\beta^{(2)}_{n+1}(\bm{\lambda}+2\bm{\delta}^{(1)})},\\
  \tilde{\alpha}^{(4)}_n(\bm{\lambda})
  &\eqdef\alpha^{(4)}_{n,0}(\bm{\lambda}+2\bm{\delta}^{(123)})
  \frac{\beta^{(1)}_{n+2}(\bm{\lambda})}{\beta^{(1)}_{n+3}(\bm{\lambda})}
  \frac{\beta^{(2)}_{n+1}(\bm{\lambda}+2\bm{\delta}^{(1)})}
  {\beta^{(2)}_{n+2}(\bm{\lambda}+2\bm{\delta}^{(1)})}
  \frac{\beta^{(3)}_n(\bm{\lambda}+2\bm{\delta}^{(12)})}
  {\beta^{(3)}_{n+1}(\bm{\lambda}+2\bm{\delta}^{(12)})}.
  \label{talpha4n}
\end{align}
It is straightforward to check that \eqref{alphan4:AW2}--\eqref{betan:AW2}
(with \eqref{alphajnk}--\eqref{betajn} and \eqref{talpha1n}--\eqref{talpha4n})
are equal to \eqref{alphan4:AW}--\eqref{betan:AW}, respectively, by direct
calculation.

\section{Differential Relations}
\label{sec:DlR}

In this section, by the same method in \S\,\ref{sec:DR}, we present some
differential relations for the orthogonal polynomials described by ordinary
quantum mechanics (oQM).
Explicit forms of various quantities are given in \S\,\ref{sec:ExJ} and
Appendix \ref{app:ExoQM}.

As in idQM, the dynamical variables of oQM are the real coordinate $x$ and its
conjugate momentum $p$, and the Hamiltonian is
\begin{align}
  \cH&\eqdef-\frac{d^2}{dx^2}+U(x)=\cA^{\dagger}\cA,\quad
  U(x)\eqdef\Bigl(\frac{dw(x)}{dx}\Bigr)^2
  +\frac{d^2w(x)}{dx^2},\\
  \cA&\eqdef\frac{d}{dx}-\frac{dw(x)}{dx},\quad
  \cA^{\dagger}=-\frac{d}{dx}-\frac{dw(x)}{dx},
\end{align}
where $w(x)$ is the prepotential and the ground state wavefunction is
$\phi_0(x)=e^{w(x)}$.
The following five polynomials \cite{kls} are studied by the oQM formulation
\cite{os24,fbsr}:
Hermite (He),
Laguerre (L),
Jacobi (J),
Bessel (B) and
pseudo Jacobi (pJ).
Their sinusoidal coordinates are
\begin{equation}
  \text{He}: \eta(x)=x,
  \ \ \text{L}: \eta(x)=x^2,
  \ \ \text{J}: \eta(x)=\cos 2x,
  \ \ \text{B}: \eta(x)=e^x,
  \ \ \text{pJ}: \eta(x)=\sinh x.
\end{equation}
We have \eqref{Hphi=}--\eqref{orthorel} with the following replacement
\cite{os24}
\begin{equation}
  \eqref{wtcH}\rightarrow
  \ \wtcH=-\frac{d^2}{dx^2}-2\frac{dw(x)}{dx}\frac{d}{dx}
  \ \Bigl(=-4c_2(\eta)\frac{d^2}{d\eta^2}-4c_1(\eta)\frac{d}{d\eta}\Bigr).
\end{equation}
Note that $c_1(\eta)$ may depend on $\bm{\lambda}$ ($c_1(\eta;\bm{\lambda})$)
but $c_2(\eta)$ does not depend on $\bm{\lambda}$
(please do not confuse these $c_j(\eta)$ ($j=1,2$) and $c_n(\bm{\lambda})$ in
\eqref{cn}).
These oQM systems also satisfy the shape invariance condition
\eqref{shapeinv} with $\kappa=1$.
We have \eqref{Aphi=}--\eqref{ImA} with the following replacements
\cite{os24,fbsr},
\begin{align}
  &\eqref{F}\rightarrow
  \ \cF=
  \constF\Bigl(\frac{d\eta(x)}{dx}\Bigr)^{-1}\frac{d}{dx}
  \ \Bigl(=\constF\frac{d}{d\eta}\Bigr),
  \label{FoQM}\\
  &\eqref{B}\rightarrow
  \ \cB(\bm{\lambda})=
  -\constF^{-1}\Bigl(\frac{d\eta(x)}{dx}\frac{d}{dx}
  +4c_1\bigl(\eta(x);\bm{\lambda}\bigr)\Bigr)
  \ \Bigl(=-4\constF^{-1}\Bigl(c_2(\eta)\frac{d}{d\eta}
  +c_1(\eta;\bm{\lambda})\Bigr)\Bigr),
  \label{BoQM}
\end{align}
where $\constF$ is a constant.
Note that $\cF$ does not depend on $\bm{\lambda}$.

Let us define a function $\check{\Phi}(x)$ as follows:
\begin{equation}
  \check{\Phi}(x)\eqdef\frac{4}{\constF^2}c_2\bigl(\eta(x)\bigr).
  \label{cPhidefoQM}
\end{equation}
Note that $\check{\Phi}(x)$ does not depend on $\bm{\lambda}$.
By using explicit forms of $c_2(\eta)$ and $\constF$,
a direct calculation shows the following proposition.
\begin{prop}\label{prop:PhioQM}
The function $\check{\Phi}(x)$ \eqref{cPhidefoQM} is a polynomial of
degree $m$ in $\eta=\eta(x)$, \eqref{cPhi},
\begin{equation}
  m=\left\{
  \begin{array}{ll}
  2&:\text{\rm J,\,B,\,pJ}\\
  1&:\text{\rm L}
  \end{array}\right..
\end{equation}
\end{prop}
\remark\label{rem:He}
For He, $\check{\Phi}(x)$ \eqref{cPhidefoQM} becomes a constant,
$\check{\Phi}(x)=1$.

This $\check{\Phi}(x)$ relates $\phi_0(x)$ with different
parameters in the following way.
\begin{prop}\label{prop:Phiphi02oQM}
$\check{\Phi}(x)$ and $\phi_0(x)$ satisfy the following relation:
\begin{equation}
  \check{\Phi}(x)\phi_0(x;\bm{\lambda})^2
  =\phi_0(x;\bm{\lambda}+\bm{\delta})^2.
  \label{Phiphi02=oQM}
\end{equation}
\end{prop}
Proof: By using explicit forms of $\check{\Phi}(x)$ and
$\phi_0(x;\bm{\lambda})=e^{w(x;\bm{\lambda})}$, a direct calculation shows
\eqref{Phiphi02=oQM}.
\hfill\fbox{}

\remark\label{rem:aaHe}
The shift of $\bm{\lambda}$ in \eqref{Phiphi02=oQM} is $\bm{\delta}$,
which differs from the shift $2\bm{\delta}$ in \eqref{Phiphi02=}.

By the same argument as in \S\,\ref{sec:DR}, we obtain the following theorems.
\begin{thm}\label{thm:PhiPnoQM}
We have the following relations,
\begin{equation}
  \check{\Phi}(x)\check{P}_n(x;\bm{\lambda}+\bm{\delta})
  =\beta_n(\bm{\lambda})\sum_{k=0}^m\alpha_{n,k}(\bm{\lambda})
  \check{P}_{n+k}(x;\bm{\lambda})
  \ \ (n\in\mathbb{Z}_{\geq 0}),
  \label{PhiPn=oQM}
\end{equation}
where $\alpha_{n,k}(\bm{\lambda})$ is given by \eqref{alphank} and
$\beta_n(\bm{\lambda})$ is given by
\begin{equation}
  \beta_n(\bm{\lambda})\eqdefrm c^{\Phi}
  \frac{c_n(\bm{\lambda}+\bm{\delta})}{c_{n+m}(\bm{\lambda})}.
  \label{betanoQM}
\end{equation}
Explicit expressions of $\check{\Phi}(x)$,
$\alpha_{n,k}(\bm{\lambda})$ and $\beta_n(\bm{\lambda})$ are given in
\S\,\ref{sec:ExJ} and Appendix \ref{app:ExoQM}.
\end{thm}
%
\begin{thm}\label{thm:bibunrel}
We have the following differential relations for $P_n(\eta;\bm{\lambda})$'s,
\begin{equation}
  \frac{4}{\constF}c_2(\eta)\frac{d}{d\eta}
  P_{n+1}(\eta;\bm{\lambda})
  =\beta^{\cF}_n(\bm{\lambda})\sum_{k=0}^m\alpha_{n,k}(\bm{\lambda})
  P_{n+k}(\eta;\bm{\lambda})
  \ \ (n\in\mathbb{Z}_{\geq 0}),
  \label{bibunrel}
\end{equation}
where $\beta^{\cF}_n(\bm{\lambda})$ is given by
\begin{equation}
  \beta^{\cF}_n(\bm{\lambda})\eqdefrm
  f_{n+1}(\bm{\lambda})\beta_n(\bm{\lambda}).
  \label{betaFnoQM}
\end{equation}
\end{thm}
%
\begin{thm}\label{thm:mapoQM}
Multiplying by $\sqrt{\check{\Phi}(x)}$ is a surjective map
from $\mathsf{H}_{\bm{\lambda}+\bm{\delta}}$ to $\mathsf{H}_{\bm{\lambda}}$,
\begin{equation}
  \sqrt{\check{\Phi}(x)}:
  \mathsf{H}_{\bm{\lambda}+\bm{\delta}}\to\mathsf{H}_{\bm{\lambda}},\quad
  \text{\rm Im}\,\sqrt{\check{\Phi}(x)}
  =\mathsf{H}_{\bm{\lambda}}.
  \label{srPhioQM}
\end{equation}
\end{thm}
\remark\label{rem:rangeoQM}
This Theorem\,\ref{thm:mapoQM} holds only for physical $\bm{\lambda}$.
On the other hand,
\eqref{FP=}--\eqref{BP=} with \eqref{FoQM}--\eqref{BoQM},
Theorem\,\ref{thm:PhiPnoQM} and Theorem\,\ref{thm:bibunrel}
hold for any $\bm{\lambda}$, because they are algebraic relations.

\subsection{Explicit expression : J case}
\label{sec:ExJ}

As an illustration, we present explicit expressions of various quantities
for J case. For other cases, see Appendix\,\ref{app:ExoQM}.

A set of parameters $\bm{\lambda}$ is
\begin{equation}
  \bm{\lambda}=(g,h)\ \ (g,h\in\mathbb{R}),
\end{equation}
and the standard parametrization \cite{kls} is
$(\alpha,\beta)^{\text{standard}}=(g-\tfrac12,h-\tfrac12)$.
The basic data of the oQM system of J type are as follows \cite{os24,fbsr}:
\begin{align}
  &x_{\text{min}}=0,\quad x_{\text{max}}=\frac{\pi}{2},\\
  &\eta(x)=\cos 2x,\quad\kappa=1,\quad\bm{\delta}=(1,1),\\
  &w(x;\bm{\lambda})=g\log\sin x+h\log\cos x,\quad
  U(x;\bm{\lambda})=\frac{g(g-1)}{\sin^2x}+\frac{h(h-1)}{\cos^2x}-(g+h)^2,\\
  &c_1(\eta;\bm{\lambda})=h-g-(g+h+1)\eta,
  \ \ c_2(\eta)=1-\eta^2,\ \ \constF=-4,\\
  &\mathcal{E}_n(\bm{\lambda})=4n(n+g+h),\quad
  f_n(\bm{\lambda})=-2(n+g+h),\quad b_n(\bm{\lambda})=-2(n+1),\\
  &\phi_0(x;\bm{\lambda})^2
  =(\sin x)^{2g}(\cos x)^{2h}
  =2^{-g-h}\bigl(1-\eta(x)\bigr)^g\bigl(1+\eta(x)\bigr)^h,\\
  &h_n(\bm{\lambda})
  =\frac{\Gamma(n+g+\frac12)\Gamma(n+h+\frac12)}
  {2\,n!\,(2n+g+h)\Gamma(n+g+h)},\\
  &P_n(\eta;\bm{\lambda})
  =P^{(g-\frac12,h-\frac12)}(\eta)
  =\frac{(g+\frac12)_n}{n!}\,{}_2F_1\Bigl(
  \genfrac{}{}{0pt}{}{-n,\,n+g+h}{g+\frac12}\Bigm|\frac{1-\eta}{2}\Bigr),
  \label{cPn:J}\\
  &c_n(\bm{\lambda})=\frac{(n+g+h)_n}{2^nn!},
\end{align}
where $P^{(\alpha,\beta)}_n(\eta)$ is the Jacobi polynomial \cite{kls}.
The physical $\bm{\lambda}$ is $g,h>\frac12$.

{}From \eqref{cPhidefoQM}, we have
\begin{equation}
  \check{\Phi}(x)=\frac14\bigl(1-\eta(x)^2\bigr)
  =(\sin x\cos x)^2,\quad
  m=2,\quad c^{\Phi}=-\frac14,
\end{equation}
and the zeros $\eta_j$ and $x_j$ ($j=1,2$) are
\begin{equation}
  \eta_j=
  \begin{cases}
  1&\!\!(j=1)\\
  -1&\!\!(j=2)
  \end{cases},\quad
  x_j=
  \begin{cases}
  0&\!\!(j=1)\\
  \frac{\pi}{2}&\!\!(j=2)
  \end{cases}.
\end{equation}
The expression \eqref{cPn:J} and the parity property
$P_n\bigl(-\eta;(g,h)\bigr)=(-1)^nP_n\bigl(\eta;(h,g)\bigr)$ give
\begin{equation}
  P_n(\eta_1;\bm{\lambda})=\frac{(g+\frac12)_n}{n!},\quad
  P_n(\eta_2;\bm{\lambda})=(-1)^n\frac{(h+\frac12)_n}{n!}.
  \label{cPnxj:J}
\end{equation}
By direct calculation, we can obtain
$D_n^{(0,1,\ldots,\breve{k},\ldots,2)}(\bm{\lambda})$ ($k=0,1,2$),
\eqref{Dndef}.
Here we write down $D_n^{(0,1)}(\bm{\lambda})$ and
$\alpha_{n,k}(\bm{\lambda})$ ($k=0,1,2$) \eqref{alphank},
\begin{align}
  &D_n^{(0,1)}(\bm{\lambda})
  =-\frac{2n+g+h+1}{n+1},\\
  &\alpha_{n,2}(\bm{\lambda})=1,
  \label{alphan2:J}\\
  &\alpha_{n,1}(\bm{\lambda})=-\frac{(g-h)(2n+g+h+2)}{(n+2)(2n+g+h+1)},\\
  &\alpha_{n,0}(\bm{\lambda})
  =-\frac{(n+g+\frac12)(n+h+\frac12)(2n+g+h+3)}{(n+1)_2(2n+g+h+1)}.
  \label{alphan0:J}
\end{align}
The constant $\beta_n(\bm{\lambda})$ \eqref{betanoQM} and
$\beta^{\cF}_n(\bm{\lambda})$ \eqref{betaFnoQM} are
\begin{equation}
  \beta_n(\bm{\lambda})=-\frac{(n+1)_2}{(2n+g+h+2)_2},\quad
  \beta^{\cF}_n(\bm{\lambda})
  =\frac{2(n+1)_2(n+g+h+1)}{(2n+g+h+2)_2}.
  \label{betan:J}
\end{equation}

\section{Summary and Comments}
\label{sec:summary}

The quantum mechanical formulation is a useful tool to study the orthogonal
polynomials in the Askey scheme.
The quantum mechanical systems associated with the orthogonal polynomials
in the Askey scheme have the shape invariance property, which relates the
Hilbert spaces $\mathsf{H}_{\bm{\lambda}}$ and
$\mathsf{H}_{\bm{\lambda}+\bm{\delta}}$, and as its consequence the forward
and backward shift relations for the orthogonal polynomials hold.
For idQM systems, the function $\check{\Phi}(x;\bm{\lambda})$ \eqref{cPhidef}
satisfies Proposition\,\ref{prop:Phi} and \ref{prop:Phiphi02}.
Then the Christoffel's theorem gives Theorem\,\ref{thm:PhiPn} and their explicit
expressions are given in \S\,\ref{sec:ExAW} and Appendix \ref{app:ExidQM}.
By combining it with the forward shift relations, we obtain the difference
relations for the orthogonal polynomials, Theorem\,\ref{thm:sabunrel}.
The function $\check{\Phi}(x;\bm{\lambda})$ relates the Hilbert spaces
$\mathsf{H}_{\bm{\lambda}}$ and $\mathsf{H}_{\bm{\lambda}+2\bm{\delta}}$,
Theorem\,\ref{thm:map}.
For oQM systems, we obtain the differential relations for the orthogonal
polynomials, Theorem\,\ref{thm:bibunrel}, and their explicit
expressions are given in \S\,\ref{sec:ExJ} and Appendix \ref{app:ExoQM}.
The function $\check{\Phi}(x)$ \eqref{cPhidefoQM} relates the Hilbert spaces
$\mathsf{H}_{\bm{\lambda}}$ and $\mathsf{H}_{\bm{\lambda}+\bm{\delta}}$,
Theorem\,\ref{thm:mapoQM}.

For oQM (L, J, B and pJ), each explicit forms of Theorem\,\ref{thm:PhiPnoQM}
and \ref{thm:bibunrel} are not so complicated. So such specific expressions
may exist in the literature, but we think that the universal expression is new.
For idQM, an explicit form of Theorem\,\ref{thm:PhiPn} for MP case is given
in \cite{is25}, and one $a_j$ shift is considered for W case in \cite{jsj14}.
But we think that Theorem\,\ref{thm:PhiPn} (for most cases), its universal
expression and Theorem\,\ref{thm:sabunrel} are new results.
These new results are obtained based on the Christoffel's theorem
(Theorem\,\ref{thm:Chr}), the shape invariance and
Proposition\,\ref{prop:Phiphi02} or \ref{prop:Phiphi02oQM}.
The shape invariance of the quantum mechanical systems (oQM, idQM and rdQM)
gives the forward and backward shift relations for all orthogonal polynomials
in the Askey scheme. To obtain Theorem\,\ref{thm:PhiPn} and \ref{thm:sabunrel}
for idQM (Theorem\,\ref{thm:PhiPnoQM} and \ref{thm:bibunrel} for oQM), we need
Proposition\,\ref{prop:Phiphi02} (Proposition\,\ref{prop:Phiphi02oQM}).
However, for rdQM, which describes orthogonal polynomials of a discrete
variable such as $q$-Racah polynomial, we do not have the property like
Proposition\,\ref{prop:Phiphi02} or \ref{prop:Phiphi02oQM} related to the
shape invariance. So we do not have new results for rdQM.

Another type of forward and backward shift relations are discussed in
\cite{fbsr}. For example, for AW case, the ratio
$\phi_0(x;\bm{\lambda}-2\bar{\bm{\delta}})^2/\phi_0(x;\bm{\lambda})^2$ has
a form $(a+b\eta(x))(c+d\eta(x))/((a'+b'\eta(x))(c'+d'\eta(x))$,
namely it is a rational function of $\eta(x)$.
So we can apply Uvarov's theorem, see Remark\,\ref{rem:Uvarov}.
It is an interesting problem to write down the explicit expressions and
to combine them with another type of forward shift relations.

The exceptional and multi-indexed orthogonal polynomials are a new type of
orthogonal polynomials \cite{gkm08,os25,os26,os27}.
They satisfy second order differential or difference equations and form a
complete set of orthogonal basis in spite of the missing degrees,
by which the restrictions of the Bochner's theorem and its generalizations
\cite{ismail,kls} are avoided.
The deformed quantum systems describing the case-(1) multi-indexed orthogonal
polynomials have the shape invariance property and the case-(1) multi-indexed
orthogonal polynomials satisfy the forward and backward shift relations.
Here ``case-(1)'' means that the set of missing degrees is
$\{0,1,\ldots,\ell-1\}$, where $\ell$ is a positive integer.
The case-(1) multi-indexed orthogonal polynomials
$\check{P}_{\cD,n}(x;\bm{\lambda})$ of a continuous variable are constructed
for oQM: Jacobi and Laguerre types \cite{os25}, and for idQM: Askey-Wilson,
Wilson \cite{os27}, continuous Hahn and Meixner-Pollaczek types \cite{idQMcH}.
The weight function for $\check{P}_{\cD,n}(x;\bm{\lambda})$ is
$\psi_{\cD}(x;\bm{\lambda})^2$ and this $\psi_{\cD}(x;\bm{\lambda})$ is
expressed in terms of $\phi_0(x;\bm{\lambda})$ (with shifted parameters) and
the denominator polynomial $\check{\Xi}_{\cD}(x;\bm{\lambda})$.
The ratio of $\psi_{\cD}(x;\bm{\lambda}+2\bm{\delta})^2$ and
$\psi_{\cD}(x;\bm{\lambda})^2$ for idQM and that of
$\psi_{\cD}(x;\bm{\lambda}+\bm{\delta})^2$ and $\psi_{\cD}(x;\bm{\lambda})^2$
for oQM are rational functions of $\eta(x)$.
Hence we can apply Uvarov's theorem (Remark\,\ref{rem:Uvarov}).
Then, by combining it with the forward shift relation, we obtain some
difference or differential relations for $\check{P}_{\cD,n}(x;\bm{\lambda})$.
It is an challenging problem to obtain their explicit expressions.

\section*{Acknowledgements}

This work is supported by JSPS KAKENHI Grant Number JP25K07147.
I thank to my late mother Kazuko Odake for warm encouragement.

\bigskip
\appendix
\section{Explicit Expressions : idQM case}
\label{app:ExidQM}

In this appendix, data for the orthogonal polynomials described by idQM are
provided. We present basic data for the polynomials \cite{os13}
and results data for \S\,\ref{sec:DR}.
Data for AW case is given in \S\,\ref{sec:ExAW}.

We remark that the standard parametrizations \cite{kls} are
\begin{align}
  \text{cH}&:(a,b,c,d)^{\text{standard}}=(a_1,a_2,a_1^*,a_2^*),\quad
  \text{MP}:(\lambda,\phi)^{\text{standard}}=(a,\phi),\n
  \text{W,\,AW}&:(a,b,c,d)^{\text{standard}}=(a_1,a_2,a_3,a_4),\quad
  \text{cdH,\,cd$q$H}:(a,b,c)^{\text{standard}}=(a_1,a_2,a_3),\n
  \text{ASC}&:(a,b)^{\text{standard}}=(a_1,a_2),\quad
  \text{c$q$H}:(a,b,c,d,\phi)^{\text{standard}}=(a_1,a_2,a_1,a_2,\phi),
\end{align}
and some polynomials are symmetric under the permutations of the following
parameters:
\begin{equation}
  \text{W,\,AW}:(a_1,a_2,a_3,a_4),\quad
  \text{cdH,\,cd$q$H}:(a_1,a_2,a_3),\quad
  \text{cH,\,ASC,\,c$q$H}:(a_1,a_2).
\end{equation}

\subsection{(\romannumeral1) $\bm{\eta(x)=x}$}
\label{app:(i)}

\begin{equation}
  x_{\text{min}}=-\infty,\quad x_{\text{max}}=\infty,\quad\gamma=1,\quad
  \eta(x)=x,\quad\varphi(x)=1,\quad\kappa=1.
\end{equation}

\subsubsection{continuous Hahn (cH)}
\label{app:cH}

\noindent
\underline{Basic data} :
\begin{align}
  &\bm{\lambda}=(a_1,a_2)\ \ (a_j\in\mathbb{C}),\quad
  \bm{\delta}=(\tfrac12,\tfrac12),\quad
  (a_3,a_4)\eqdef(a_1^*,a_2^*),\quad b_j:\eqref{bj},\\
  &V(x;\bm{\lambda})=(a_1+ix)(a_2+ix),\\
  &\mathcal{E}_n(\bm{\lambda})=n(n+b_1-1),\quad
  f_n(\bm{\lambda})=n+b_1-1,\quad
  b_n(\bm{\lambda})=n+1,\\
  &\phi_0(x;\bm{\lambda})^2
  =\prod_{j=1}^2\Gamma(a_j+ix)\cdot\prod_{k=3}^4\Gamma(a_k-ix),\quad
  h_n(\bm{\lambda})
  =\frac{2\pi\prod_{j=1}^2\prod_{k=3}^4\Gamma(n+a_j+a_k)}
  {n!(2n+b_1-1)\Gamma(n+b_1-1)},\\
  &\check{P}_n(x;\bm{\lambda})=p_n(x;a_1,a_2,a_3,a_4)\n
  &\phantom{\check{P}_n(x;\bm{\lambda})}
  =\frac{i^n(a_1+a_3,a_1+a_4)_n}{n!}
  {}_3F_2\Bigl(\genfrac{}{}{0pt}{}{-n,\,n+b_1-1,\,a_1+ix}
  {a_1+a_3,\,a_1+a_4}\Bigm|1\Bigr),\\
  &c_n(\bm{\lambda})=\frac{(n+b_1-1)_n}{n!},
\end{align}
where $p_n(x;a_1,a_2,a_3,a_4)$ is the continuous Hahn polynomial \cite{kls}.
The physical $\bm{\lambda}$ is $\text{Re}\,a_j>0$ ($j=1,2$).

\noindent
\underline{Results data} :
\begin{align}
  &\check{\Phi}(x;\bm{\lambda})=\prod_{j=1}^2(a_j+ix)\cdot
  \prod_{j=3}^4(a_j-ix),\quad
  m=4,\quad c^{\Phi}(\bm{\lambda})=1,\\
  &\eta_j=x_j=
  \begin{cases}
  ia_j&\!\!(j=1,2)\\
  -ia_j&\!\!(j=3,4)
  \end{cases},\quad
  \check{P}_n(x_j;\bm{\lambda})
  =\begin{cases}
  {\displaystyle\frac{i^n}{n!}\prod_{k=3}^4(a_j+a_k)_n}&\!\!(j=1,2)\\
  {\displaystyle\frac{(-i)^n}{n!}\prod_{k=1}^2(a_j+a_k)_n}&\!\!(j=3,4)
  \end{cases},\\
  &D_n^{(0,1,2,3)}(\bm{\lambda})
  =(a_1-a_2)(a_3-a_4)\prod_{j=1}^2\prod_{k=3}^4(a_j+a_k+n)\cdot
  \frac{(b_1+2n)_3(b_1+2n+2)_3}{(n+1)_3(n+1)_2(n+1)},\\
  &\alpha_{n,4}(\bm{\lambda})=1,\\
  &\alpha_{n,3}(\bm{\lambda})=-i(a_1-a_2+a_3-a_4)(a_1-a_2-a_3+a_4)
  \frac{b_1+2n+5}{(n+4)(b_1+2n+2)},\\
  &\alpha_{n,2}(\bm{\lambda})=
  \frac{(b_1+2n+3)(b_1+2n+6)}{(n+3)_2(b_1+2n+1)_2}
  \Bigl((2n+3)(b_1+n)_3-(n+1)(2n+1)(b_1+n+1)_2\n
  &\phantom{\alpha_{n,2}(\bm{\lambda})}\quad
  +(n+1)_2(2n+3)(b_1+n+2)-(n+1)_3
  -6b_2^2+b_3(3b_1+4n+6)+4b_4\n
  &\phantom{\alpha_{n,2}(\bm{\lambda})}\quad
  +2\bigl((b_1+n)_2-4(n+1)(b_1+n+1)+n(n+1)\bigr)b_2\n
  &\phantom{\alpha_{n,2}(\bm{\lambda})}\quad
  -(a_1^3+a_2^3)(a_3+a_4)-(a_1+a_2)(a_3^3+a_4^3)
  -(a_1^2+a_2^2)a_3a_4-a_1a_2(a_3^2+a_4^2)\n
  &\phantom{\alpha_{n,2}(\bm{\lambda})}\quad
  +4(a_1^2+a_2^2)(a_3^2+a_4^2)
  +(2n+3)\bigl((a_1^2+a_2^2)(a_3+a_4)+(a_1+a_2)(a_3^2+a_4^2)\bigr)\n
  &\phantom{\alpha_{n,2}(\bm{\lambda})}\quad
  +2(n+1)_2(a_1+a_2)(a_3+a_4)\Bigr),\\
  &\alpha_{n,1}(\bm{\lambda})=i(a_1-a_2+a_3-a_4)(a_1-a_2-a_3+a_4)
  \frac{(b_1+2n+5)_2}{(n+2)_3(b_1+2n)(b_1+2n+2)}\n
  &\phantom{\alpha_{n,1}(\bm{\lambda})=}\times
  \prod_{j=1}^2\prod_{k=3}^4(a_j+a_k+n+1),\\
  &\alpha_{n,0}(\bm{\lambda})=\prod_{j=1}^2\prod_{k=3}^4(a_j+a_k+n)_2\cdot
  \frac{(b_1+2n+4)_3}{(n+1)_4(b_1+2n)_3},\\
  &\beta_n(\bm{\lambda})=\frac{(n+1)_4}{(b_1+2n+3)_4},\quad
  \beta^{\cF}_n(\bm{\lambda})=\frac{(n+1)_4(b_1+n+1)_2}{(b_1+2n+3)_4}.
\end{align}

\subsubsection{Meixner-Pollaczek (MP)}
\label{app:MP}

\noindent
\underline{Basic data} :
\begin{align}
  &\bm{\lambda}=(a,\phi)\ \ (a,\phi\in\mathbb{R}),\quad
  \bm{\delta}=(\tfrac12,0),\\
  &V(x;\bm{\lambda})=e^{i(\frac{\pi}{2}-\phi)}(a+ix),\\
  &\mathcal{E}_n(\bm{\lambda})=2n\sin\phi,\quad
  f_n(\bm{\lambda})=2\sin\phi,\quad
  b_n(\bm{\lambda})=n+1,\\
  &\phi_0(x;\bm{\lambda})^2
  =e^{(2\phi-\pi)x}\Gamma(a+ix)\Gamma(a-ix),\quad
  h_n(\bm{\lambda})
  =\frac{2\pi\,\Gamma(n+2a)}{n!\,(2\sin\phi)^{2a}},\\
  &\check{P}_n(x;\bm{\lambda})=P^{(a)}_n(x;\phi)
  =\frac{(2a)_ne^{in\phi}}{n!}
  {}_2F_1\Bigl(
  \genfrac{}{}{0pt}{}{-n,\,a+ix}{2a}\Bigm|1-e^{-2i\phi}\Bigr),\\
  &c_n(\bm{\lambda})=\frac{(2\sin\phi)^n}{n!},
\end{align}
where $P^{(a)}_n(x;\phi)$ is the Meixner-Pollaczek polynomial \cite{kls}.
The physical $\bm{\lambda}$ is $a>0$ and $0<\phi<\pi$.

\noindent
\underline{Results data} :
\begin{align}
  &\check{\Phi}(x;\bm{\lambda})=(a+ix)(a-ix),\quad
  m=2,\quad c^{\Phi}(\bm{\lambda})=1,\\
  &\eta_j=x_j=
  \begin{cases}
  ia&\!\!(j=1)\\
  -ia&\!\!(j=2)
  \end{cases},\quad
  \check{P}_n(x_j;\bm{\lambda})
  =\begin{cases}
  {\displaystyle\frac{(2a)_ne^{in\phi}}{n!}}&\!\!(j=1)\\[4pt]
  {\displaystyle\frac{(2a)_ne^{-in\phi}}{n!}}&\!\!(j=2)
  \end{cases},\\
  &D_n^{(0,1)}(\bm{\lambda})
  =-2i\sin\phi\frac{2a+n}{n+1},\\
  &\alpha_{n,2}(\bm{\lambda})=1,\quad
  \alpha_{n,1}(\bm{\lambda})=-2\cos\phi\frac{2a+n+1}{n+2},\quad
  \alpha_{n,0}(\bm{\lambda})=\frac{(2a+n)_2}{(n+1)_2},\\
  &\beta_n(\bm{\lambda})=\frac{(n+1)_2}{(2\sin\phi)^2},\quad
  \beta^{\cF}_n(\bm{\lambda})=(n+1)_2.
\end{align}

\subsection{(\romannumeral2) $\bm{\eta(x)=x^2}$}
\label{app:(ii)}

\begin{equation}
  x_{\text{min}}=0,\quad x_{\text{max}}=\infty,\quad\gamma=1,\quad
  \eta(x)=x^2,\quad\varphi(x)=2x,\quad\kappa=1.
\end{equation}

\subsubsection{Wilson (W)}
\label{app:W}

\noindent
\underline{Basic data} :
\begin{align}
  &\bm{\lambda}=(a_1,a_2,a_3,a_4)\ \ (a_j\in\mathbb{C}),\quad
  \{a_1^*,a_2^*,a_3^*,a_4^*\}=\{a_1,a_2,a_3,a_4\}\ (\text{as a set}),\n
  &\qquad\bm{\delta}=(\tfrac12,\tfrac12,\tfrac12,\tfrac12),\quad
  b_j:\eqref{bj},\\
  &V(x;\bm{\lambda})=\frac{(a_1+ix)(a_2+ix)(a_3+ix)(a_4+ix)}{2ix(2ix+1)},\\
  &\mathcal{E}_n(\bm{\lambda})=n(n+b_1-1),\quad
  f_n(\bm{\lambda})=-n(n+b_1-1),\quad
  b_n(\bm{\lambda})=-1,\\
  &\phi_0(x;\bm{\lambda})^2
  =\frac{\prod_{j=1}^4\Gamma(a_j+ix)\Gamma(a_j-ix)}{\Gamma(2ix)\Gamma(-2ix)},\\
  &h_n(\bm{\lambda})
  =\frac{2\pi n!\,(n+b_1-1)_n\prod_{1\leq j<k\leq 4}\Gamma(n+a_j+a_k)}
  {\Gamma(2n+b_1)},\\
  &\check{P}_n(x;\bm{\lambda})=W_n(x^2;a_1,a_2,a_3,a_4)\n
  &\phantom{\check{P}_n(x;\bm{\lambda})}
  =(a_1+a_2,a_1+a_3,a_1+a_4)_n\,
  {}_4F_3\Bigl(\genfrac{}{}{0pt}{}{-n,\,n+b_1-1,\,a_1+ix,\,a_1-ix}
  {a_1+a_2,\,a_1+a_3,\,a_1+a_4}\Bigm|1\Bigr),\\
  &c_n(\bm{\lambda})=(-1)^n(n+b_1-1)_n,
\end{align}
where $W_n(\eta;a_1,a_2,a_3,a_4)$ is the Wilson polynomial \cite{kls}.
The physical $\bm{\lambda}$ is $\text{Re}\,a_j>0$ ($j=1,2,3,4$).

\noindent
\underline{Results data} :
\begin{align}
  &\check{\Phi}(x;\bm{\lambda})=\prod_{j=1}^4(a_j+ix)(a_j-ix),\quad
  m=4,\quad c^{\Phi}(\bm{\lambda})=1,\\
  &\eta_j=-a_j^2,\quad
  x_j=ia_j\ \ (j=1,2,3,4),\\
  &\check{P}_n(x_j;\bm{\lambda})
  =\prod_{\substack{k=1\\k\neq j}}^4(a_j+a_k)_n
  =\frac{\prod_{k=1}^4(a_j+a_k)_n}{(2a_j)_n}
  \ \ (j=1,2,3,4),\\
  &D_n^{(0,1,2,3)}(\bm{\lambda})
  =\prod_{1\leq j<k\leq 4}(a_j-a_k)(a_j+a_k+n)\cdot
  (b_1+2n)_3(b_1+2n+2)_3,\\
  &\alpha_{n,4}(\bm{\lambda})=1,\\
  &\alpha_{n,3}(\bm{\lambda})=-\frac{b_1+2n+5}{b_1+2n+2}\bigl(
  (b_1+n+1)_3+(3n^2+12n+11)b_1+3(n+1)_3-2b_1b_2+4b_3\bigr),\\
  &\alpha_{n,2}(\bm{\lambda})=
  \frac{(b_1+2n+3)(b_1+2n+6)}{(b_1+2n+1)_2}
  \Bigl((3n^2+9n+7)(b_1+n)_4\n
  &\phantom{\alpha_{n,2}(\bm{\lambda})}\quad
  +\bigl(b_3+(2n+3)(2b_2-1)\bigr)(b_1+n)_3\n
  &\phantom{\alpha_{n,2}(\bm{\lambda})}\quad
  +\bigl(2b_4-(5n+6)b_3+b_2(b_2-2(n+1)(2n+1))\n
  &\phantom{\alpha_{n,2}(\bm{\lambda})}\qquad
  +3(n+1)(n^3+5n^2+10n+7)\bigr)(b_1+n)_2\n
  &\phantom{\alpha_{n,2}(\bm{\lambda})}\quad
  +\bigl(2(2n+5)b_4+(5n(n+1)-6b_2)b_3-2(n+1)b_2(2b_2-2n^2-3n-4)\n
  &\phantom{\alpha_{n,2}(\bm{\lambda})}\qquad
  +(n+1)(6n^3+28n^2+53n+36)\bigr)(b_1+n)\n
  &\phantom{\alpha_{n,2}(\bm{\lambda})}\quad
  +2(n+2)(n+4)b_4+b_3\bigl(6b_3+6nb_2-n(n+1)(n-4)\bigr)\\
  &\phantom{\alpha_{n,2}(\bm{\lambda})}\quad
  +(n+1)b_2\bigl((n-4)b_2+2(3n^2+5n+4)\bigr)
  +(n+1)(7n^3+32n^2+57n+36)\Bigr),\n
  &\alpha_{n,1}(\bm{\lambda})=-\prod_{1\leq j<k\leq 4}(a_j+a_k+n+1)\cdot
  \frac{(b_1+2n+5)_2}{(b_1+2n)(b_1+2n+2)}\n
  &\phantom{\alpha_{n,1}(\bm{\lambda})=}\times
  \bigl(3(n+1)(b_1+n)_2+(3n+1)b_1+n(n+1)(n+5)+2b_1b_2-4b_3\bigr),\\
  &\alpha_{n,0}(\bm{\lambda})=\prod_{1\leq j<k\leq 4}(a_j+a_k+n)_2\cdot
  \frac{(b_1+2n+4)_3}{(b_1+2n)_3},\\
  &\beta_n(\bm{\lambda})=\frac{1}{(b_1+2n+3)_4},\quad
  \beta^{\cF}_n(\bm{\lambda})=\frac{(n+1,b_1+n+1)_2}{(b_1+2n+3)_4}.
\end{align}

\subsubsection{continuous dual Hahn (cdH)}
\label{app:cdH}

\noindent
\underline{Basic data} :
\begin{align}
  &\bm{\lambda}=(a_1,a_2,a_3)\ \ (a_j\in\mathbb{C}),\quad
  \{a_1^*,a_2^*,a_3^*\}=\{a_1,a_2,a_3\}\ (\text{as a set}),\quad
  \bm{\delta}=(\tfrac12,\tfrac12,\tfrac12),\\
  &V(x;\bm{\lambda})=\frac{(a_1+ix)(a_2+ix)(a_3+ix)}{2ix(2ix+1)},\\
  &\mathcal{E}_n(\bm{\lambda})=n,\quad
  f_n(\bm{\lambda})=-n,\quad
  b_n(\bm{\lambda})=-1,\\
  &\phi_0(x;\bm{\lambda})^2
  =\frac{\prod_{j=1}^3\Gamma(a_j+ix)\Gamma(a_j-ix)}{\Gamma(2ix)\Gamma(-2ix)},
  \quad
  h_n(\bm{\lambda})
  =2\pi n!\prod_{1\leq j<k\leq 3}\Gamma(n+a_j+a_k),\\
  &\check{P}_n(x;\bm{\lambda})=S_n(x^2;a_1,a_2,a_3)
  =(a_1+a_2,a_1+a_3)_n\,
  {}_3F_2\Bigl(\genfrac{}{}{0pt}{}{-n,\,a_1+ix,\,a_1-ix}
  {a_1+a_2,\,a_1+a_3}\Bigm|1\Bigr),\\
  &c_n(\bm{\lambda})=(-1)^n,
\end{align}
where $S_n(\eta;a_1,a_2,a_3)$ is the continuous dual Hahn polynomial \cite{kls}.
The physical $\bm{\lambda}$ is $\text{Re}\,a_j>0$ ($j=1,2,3$).

\noindent
\underline{Results data} :
\begin{align}
  &\check{\Phi}(x;\bm{\lambda})=\prod_{j=1}^3(a_j+ix)(a_j-ix),\quad
  m=3,\quad c^{\Phi}(\bm{\lambda})=1,\\
  &\eta_j=-a_j^2,\quad
  x_j=ia_j\ \ (j=1,2,3),\\
  &\check{P}_n(x_j;\bm{\lambda})
  =\prod_{\substack{k=1\\k\neq j}}^3(a_j+a_k)_n
  =\frac{\prod_{k=1}^3(a_j+a_k)_n}{(2a_j)_n}
  \ \ (j=1,2,3),\\
  &D_n^{(0,1,2)}(\bm{\lambda})
  =-\prod_{1\leq j<k\leq 3}(a_j-a_k)(a_j+a_k+n),\\
  &\alpha_{n,3}(\bm{\lambda})=1,\\
  &\alpha_{n,2}(\bm{\lambda})=-\bigl(b_2+(b_1+n)_2+(2n+5)(b_1+n+1)+n+2\bigr),\\
  &\alpha_{n,1}(\bm{\lambda})=
  \prod_{1\leq j<k\leq 3}(a_j+a_k+n+1)\cdot\bigl(2b_1+3(n+1)\bigr),\\
  &\alpha_{n,0}(\bm{\lambda})=
  -\prod_{1\leq j<k\leq 3}(a_j+a_k+n)_2,\\
  &\beta_n(\bm{\lambda})=-1,\quad
  \beta^{\cF}_n(\bm{\lambda})=-(n+1)_2,
\end{align}
where $b_1=a_1+a_2+a_3$ and $b_2=a_1a_2+a_1a_3+a_2a_3$.

\subsection{(\romannumeral3) $\bm{\eta(x)=\cos x}$}
\label{app:(iii)}

\begin{equation}
  x_{\text{min}}=0,\quad x_{\text{max}}=\pi,\quad\gamma=\log q,\quad
  \eta(x)=\cos x,\quad\varphi(x)=2\sin x,\quad\kappa=q^{-1}.
\end{equation}

\subsubsection{continuous dual $\bm{q}$-Hahn (cd$\bm{q}$H)}
\label{app:cdqH}

\noindent
\underline{Basic data} :
\begin{align}
  &q^{\bm{\lambda}}=(a_1,a_2,a_3)\ \ (a_j\in\mathbb{C}),\quad
  \{a_1^*,a_2^*,a_3^*\}=\{a_1,a_2,a_3\}\ (\text{as a set}),\quad
  \bm{\delta}=(\tfrac12,\tfrac12,\tfrac12),\\
  &V(x;\bm{\lambda})=\frac{(1-a_1e^{ix})(1-a_2e^{ix})(1-a_3e^{ix})}
  {(1-e^{2ix})(1-qe^{2ix})},\\
  &\mathcal{E}_n(\bm{\lambda})=q^{-n}-1,\quad
  f_n(\bm{\lambda})=q^{\frac{n}{2}}(q^{-n}-1),\quad
  b_n(\bm{\lambda})=q^{-\frac{n+1}{2}},\\
  &\phi_0(x;\bm{\lambda})^2
  =\frac{(e^{2ix},e^{-2ix};q)_{\infty}}
  {\prod_{j=1}^3(a_je^{ix},a_je^{-ix};q)_{\infty}},\quad
  h_n(\bm{\lambda})
  =\frac{2\pi}{(q^{n+1};q)_{\infty}
  \prod_{1\leq j<k\leq 3}(a_ja_kq^n;q)_{\infty}},\\
  &\check{P}_n(x;\bm{\lambda})=p_n(\cos x;a_1,a_2,a_3|q)
  =\frac{(a_1a_2,a_1a_3\,;q)_n}{a_1^n}
  {}_3\phi_2\Bigl(\genfrac{}{}{0pt}{}{q^{-n},\,a_1e^{ix},\,a_1e^{-ix}}
  {a_1a_2,\,a_1a_3}\Bigm|q,q\Bigr),\\
  &c_n(\bm{\lambda})=2^n,
\end{align}
where $p_n(\eta;a_1,a_2,a_3|q)$ is the continuous dual $q$-Hahn polynomial
\cite{kls}.
The physical $\bm{\lambda}$ is $|a_j|<1$ ($j=1,2,3$).

\noindent
\underline{Results data} :
\begin{align}
  &\check{\Phi}(x;\bm{\lambda})=\prod_{j=1}^3(1-a_je^{ix})(1-a_je^{-ix}),\quad
  m=3,\quad c^{\Phi}(\bm{\lambda})=-8b_3,\\
  &\eta_j=\frac{a_j+a_j^{-1}}{2},\quad
  e^{ix_j}=a_j\ \ (j=1,2,3),\\
  &\check{P}_n(x_j;\bm{\lambda})
  =a_j^{-n}\prod_{\substack{k=1\\k\neq j}}^3(a_ja_k;q)_n
  =\frac{\prod_{k=1}^3(a_ja_k;q)_n}{a_j^n(a_j^2;q)_n}
  \ \ (j=1,2,3),\\
  &D_n^{(0,1,2)}(\bm{\lambda})
  =\prod_{1\leq j<k\leq 3}(a_j-a_k)(1-a_ja_kq^n)\cdot b_3^{-2},\\
  &\alpha_{n,3}(\bm{\lambda})=1,\\
  &\alpha_{n,2}(\bm{\lambda})=-b_3^{-1}\bigl(b_2-(1+q)b_1b_3q^{n+1}
  +(1+q+q^2)b_3^2q^{2n+2}\bigr),\\
  &\alpha_{n,1}(\bm{\lambda})=
  \prod_{1\leq j<k\leq 3}(1-a_ja_kq^{n+1})\cdot b_3^{-1}
  \bigl(b_1-(1+q+q^2)b_3q^n\bigr),\\
  &\alpha_{n,0}(\bm{\lambda})=
  -\prod_{1\leq j<k\leq 3}(a_ja_kq^n;q)_2\cdot b_3^{-1},\\
  &\beta_n(\bm{\lambda})=-b_3,\quad
  \beta^{\cF}_n(\bm{\lambda})=-b_3q^{-n-\frac32}(q^{n+1};q)_2,
\end{align}
where $b_1=a_1+a_2+a_3$, $b_2=a_1a_2+a_1a_3+a_2a_3$ and $b_3=a_1a_2a_3$.

\subsubsection{Al-Salam-Chihara (ASC)}
\label{app:ASC}

\noindent
\underline{Basic data} :
\begin{align}
  &q^{\bm{\lambda}}=(a_1,a_2)\ \ (a_j\in\mathbb{C}),\quad
  \{a_1^*,a_2^*\}=\{a_1,a_2\}\ (\text{as a set}),\quad
  \bm{\delta}=(\tfrac12,\tfrac12),\\
  &V(x;\bm{\lambda})=\frac{(1-a_1e^{ix})(1-a_2e^{ix})}
  {(1-e^{2ix})(1-qe^{2ix})},\\
  &\mathcal{E}_n(\bm{\lambda})=q^{-n}-1,\quad
  f_n(\bm{\lambda})=q^{\frac{n}{2}}(q^{-n}-1),\quad
  b_n(\bm{\lambda})=q^{-\frac{n+1}{2}},\\
  &\phi_0(x;\bm{\lambda})^2
  =\frac{(e^{2ix},e^{-2ix};q)_{\infty}}
  {\prod_{j=1}^2(a_je^{ix},a_je^{-ix};q)_{\infty}},\quad
  h_n(\bm{\lambda})
  =\frac{2\pi}{(q^{n+1},a_1a_2q^n;q)_{\infty}},\\
  &\check{P}_n(x;\bm{\lambda})=Q_n(\cos x;a_1,a_2|q)
  =\frac{(a_1a_2\,;q)_n}{a_1^n}
  {}_3\phi_2\Bigl(\genfrac{}{}{0pt}{}{q^{-n},\,a_1e^{ix},\,a_1e^{-ix}}
  {a_1a_2,\,0}\Bigm|q,q\Bigr),\\
  &c_n(\bm{\lambda})=2^n,
\end{align}
where $Q_n(\eta;a_1,a_2|q)$ is the Al-Salam-Chihara polynomial \cite{kls}.
The physical $\bm{\lambda}$ is $|a_j|<1$ ($j=1,2$).

\noindent
\underline{Results data} :
\begin{align}
  &\check{\Phi}(x;\bm{\lambda})=\prod_{j=1}^2(1-a_je^{ix})(1-a_je^{-ix}),\quad
  m=2,\quad c^{\Phi}(\bm{\lambda})=4b_2,\\
  &\eta_j=\frac{a_j+a_j^{-1}}{2},\quad
  e^{ix_j}=a_j\ \ (j=1,2),\\
  &\check{P}_n(x_j;\bm{\lambda})
  =a_j^{-n}\prod_{\substack{k=1\\k\neq j}}^2(a_ja_k;q)_n
  =a_j^{-n}(a_1a_2;q)_n
  \ \ (j=1,2),\\
  &D_n^{(0,1)}(\bm{\lambda})
  =\frac{(a_1-a_2)(1-a_1a_2q^n)}{a_1a_2},\\
  &\alpha_{n,2}(\bm{\lambda})=1,\quad
  \alpha_{n,1}(\bm{\lambda})=-\frac{(a_1+a_2)(1-a_1a_2q^{n+1})}{a_1a_2},\quad
  \alpha_{n,0}(\bm{\lambda})=\frac{(a_1a_2q^n;q)_2}{a_1a_2},\\
  &\beta_n(\bm{\lambda})=a_1a_2,\quad
  \beta^{\cF}_n(\bm{\lambda})=a_1a_2q^{-n-\frac32}(q^{n+1};q)_2.
\end{align}

\subsubsection{continuous big $\bm{q}$-Hermite (cb$\bm{q}$He)}
\label{app:cbqHe}

\noindent
\underline{Basic data} :
\begin{align}
  &q^{\bm{\lambda}}=a\ \ (a\in\mathbb{R}),\quad
  \bm{\delta}=\tfrac12,\\
  &V(x;\bm{\lambda})=\frac{1-ae^{ix}}{(1-e^{2ix})(1-qe^{2ix})},\\
  &\mathcal{E}_n(\bm{\lambda})=q^{-n}-1,\quad
  f_n(\bm{\lambda})=q^{\frac{n}{2}}(q^{-n}-1),\quad
  b_n(\bm{\lambda})=q^{-\frac{n+1}{2}},\\
  &\phi_0(x;\bm{\lambda})^2
  =\frac{(e^{2ix},e^{-2ix};q)_{\infty}}{(ae^{ix},ae^{-ix};q)_{\infty}},\quad
  h_n(\bm{\lambda})
  =\frac{2\pi}{(q^{n+1};q)_{\infty}},\\
  &\check{P}_n(x;\bm{\lambda})=H_n(\cos x;a|q)
  =a^{-n}{}_3\phi_2\Bigl(
  \genfrac{}{}{0pt}{}{q^{-n},\,ae^{ix},\,ae^{-ix}}
  {0,\,0}\Bigm|q,q\Bigr),\\
  &c_n(\bm{\lambda})=2^n,
\end{align}
where $H_n(\eta;a|q)$ is the continuous big $q$-Hermite polynomial \cite{kls}.
The physical $\bm{\lambda}$ is $|a|<1$.

\noindent
\underline{Results data} :
\begin{align}
  &\check{\Phi}(x;\bm{\lambda})=(1-ae^{ix})(1-ae^{-ix}),\quad
  m=1,\quad c^{\Phi}(\bm{\lambda})=-2a,\\
  &\eta_1=\frac{a+a^{-1}}{2},\quad
  e^{ix_1}=a,\quad
  \check{P}_n(x_1;\bm{\lambda})
  =a^{-n},\\
  &D_n^{(0)}(\bm{\lambda})
  =1,\\
  &\alpha_{n,1}(\bm{\lambda})=1,\quad
  \alpha_{n,0}(\bm{\lambda})=-a^{-1},\\
  &\beta_n(\bm{\lambda})=-a,\quad
  \beta^{\cF}_n(\bm{\lambda})=-aq^{-n-\frac32}(q^{n+1};q)_2.
\end{align}

\subsubsection{continuous $\bm{q}$-Hermite (c$\bm{q}$He)}
\label{app:cqHe}

\noindent
\underline{Basic data} :
\begin{align}
  &\bm{\lambda}:\text{none},\quad
  \bm{\delta}:\text{none},\\
  &V(x;\bm{\lambda})=\frac{1}{(1-e^{2ix})(1-qe^{2ix})},\\
  &\mathcal{E}_n(\bm{\lambda})=q^{-n}-1,\quad
  f_n(\bm{\lambda})=q^{\frac{n}{2}}(q^{-n}-1),\quad
  b_n(\bm{\lambda})=q^{-\frac{n+1}{2}},\\
  &\phi_0(x;\bm{\lambda})^2
  =(e^{2ix},e^{-2ix};q)_{\infty},\quad
  h_n(\bm{\lambda})
  =\frac{2\pi}{(q^{n+1};q)_{\infty}},\\
  &\check{P}_n(x;\bm{\lambda})=H_n(\cos x|q)
  =e^{inx}{}_2\phi_0\Bigl(
  \genfrac{}{}{0pt}{}{q^{-n},\,0}{-}\Bigm|q,q^ne^{-2ix}\Bigr),\\
  &c_n(\bm{\lambda})=2^n,
\end{align}
where $H_n(\eta|q)$ is the continuous $q$-Hermite polynomial \cite{kls}.

\noindent
\underline{Results data} : none

\subsubsection{continuous $\bm{q}$-Jacobi (c$\bm{q}$J)}
\label{app:cqJ}

\noindent
\underline{Basic data} :
\begin{align}
  &\bm{\lambda}=(\alpha,\beta)\ \ (\alpha,\beta\in\mathbb{R}),\quad
  \bm{\delta}=(1,1),\\
  &V(x;\bm{\lambda})=\frac{(1-q^{\frac12(\alpha+\frac12)}e^{ix})
  (1-q^{\frac12(\alpha+\frac32)}e^{ix})
  (1+q^{\frac12(\beta+\frac12)}e^{ix})
  (1+q^{\frac12(\beta+\frac32)}e^{ix})}
  {(1-e^{2ix})(1-qe^{2ix})},\\
  &\mathcal{E}_n(\bm{\lambda})=(q^{-n}-1)(1-q^{n+\alpha+\beta+1}),\quad
  f_n(\bm{\lambda})=
  \frac{q^{\frac12(\alpha+\frac32)}q^{-n}(1-q^{n+\alpha+\beta+1})}
  {(1+q^{\frac12(\alpha+\beta+1)})(1+q^{\frac12(\alpha+\beta+2)})},\n
  &b_n(\bm{\lambda})=q^{-\frac12(\alpha+\frac32)}q^{n+1}(q^{-n-1}-1)
  (1+q^{\frac12(\alpha+\beta+1)})(1+q^{\frac12(\alpha+\beta+2)}),\\
  &\phi_0(x;\bm{\lambda})^2
  =\frac{(e^{2ix},e^{-2ix};q)_{\infty}}
  {(q^{\frac12(\alpha+\frac12)}e^{ix},q^{\frac12(\alpha+\frac12)}e^{-ix},
  -q^{\frac12(\beta+\frac12)}e^{ix},-q^{\frac12(\beta+\frac12)}e^{-ix};
  q^{\frac12})_{\infty}},\\
  &h_n(\bm{\lambda})
  =2\pi\,q^{(\alpha+\frac12)n}
  \frac{(1-q^{\alpha+\beta+1})
  (q^{\alpha+1},q^{\beta+1},-q^{\frac12(\alpha+\beta+3)};q)_n}
  {(1-q^{2n+\alpha+\beta+1})
  (q,q^{\alpha+\beta+1},-q^{\frac12(\alpha+\beta+1)};q)_n}\n
  &\phantom{h_n(\bm{\lambda})}\quad\times
  \frac{(q^{\frac12(\alpha+\beta+2)},
  q^{\frac12(\alpha+\beta+3)};q)_{\infty}}
  {(q,q^{\alpha+1},q^{\beta+1},-q^{\frac12(\alpha+\beta+1)},
  -q^{\frac12(\alpha+\beta+2)};q)_{\infty}},\\
  &\check{P}_n(x;\bm{\lambda})=P^{(\alpha,\beta)}_n(\cos x|q)\n
  &\phantom{\check{P}_n(x;\bm{\lambda})}
  =\frac{(q^{\alpha+1}\,;q)_n}{(q\,;q)_n}
  {}_4\phi_3\Bigl(
  \genfrac{}{}{0pt}{}{q^{-n},\,q^{n+\alpha+\beta+1},\,
  q^{\frac12(\alpha+\frac12)}e^{ix},\,q^{\frac12(\alpha+\frac12)}e^{-ix}}
  {q^{\alpha+1},\,-q^{\frac12(\alpha+\beta+1)},\,-q^{\frac12(\alpha+\beta+2)}}
  \Bigm|q,q\Bigr),\\
  &c_n(\bm{\lambda})=\frac{2^nq^{\frac12(\alpha+\frac12)n}
  (q^{n+\alpha+\beta+1};q)_n}
  {(q,-q^{\frac12(\alpha+\beta+1)},-q^{\frac12(\alpha+\beta+2)};q)_n},
\end{align}
where $P^{(\alpha,\beta)}_n(\eta|q)$ is the continuous $q$-Jacobi polynomial
\cite{kls}.
The physical $\bm{\lambda}$ is $\alpha,\beta\geq-\frac12$.

\noindent
\underline{Results data} :
\begin{align}
  &\check{\Phi}(x;\bm{\lambda})
  =(q^{\frac12(\alpha+\frac12)}e^{ix},q^{\frac12(\alpha+\frac12)}e^{-ix},
  -q^{\frac12(\beta+\frac12)}e^{ix},-q^{\frac12(\beta+\frac12)}e^{-ix};
  q^{\frac12})_2,\n
  &\phantom{\check{\Phi}(x;\bm{\lambda})=}
  m=4,\quad c^{\Phi}(\bm{\lambda})=16q^{\alpha+\beta+2},\\
  &\eta_j=\eta(x_j),\quad
  (e^{ix_1},e^{ix_2},e^{ix_3},e^{ix_4})
  =(q^{\frac12(\alpha+\frac12)},q^{\frac12(\alpha+\frac32)},
  -q^{\frac12(\beta+\frac12)},-q^{\frac12(\beta+\frac32)}),\\
%
\ignore{
  &\check{P}_n(x_1;\bm{\lambda})=\frac{(q^{\alpha+1};q)_n}{(q;q)_n},\quad
  \check{P}_n(x_2;\bm{\lambda})
  =\frac{q^{-\frac{n}{2}}(q^{\alpha+1};q)_n}{(q;q)_n}
  \frac{1+q^{\frac12(\alpha+\beta+1)+n}}{1+q^{\frac12(\alpha+\beta+1)}},\n
  &\check{P}_n(x_3;\bm{\lambda})=\frac{(-1)^nq^{\frac12(\alpha-\beta)n}
  (q^{\beta+1};q)_n}{(q;q)_n},\n
  &\check{P}_n(x_4;\bm{\lambda})=\frac{(-1)^nq^{\frac12(\alpha-\beta-1)n}
  (q^{\beta+1};q)_n}{(q;q)_n}
  \frac{1+q^{\frac12(\alpha+\beta+1)+n}}{1+q^{\frac12(\alpha+\beta+1)}},\\
}
  &\check{P}_n(x_1;\bm{\lambda})=\frac{(q^{\alpha+1};q)_n}{(q;q)_n},\quad
  \check{P}_n(x_4;\bm{\lambda})=\frac{(-1)^nq^{\frac12(\alpha-\beta-1)n}
  (q^{\beta+1};q)_n}{(q;q)_n}
  \frac{1+q^{\frac12(\alpha+\beta+1)+n}}{1+q^{\frac12(\alpha+\beta+1)}},\n
  &\check{P}_n(x_2;\bm{\lambda})
  =\frac{q^{-\frac{n}{2}}(q^{\alpha+1};q)_n}{(q;q)_n}
  \frac{1+q^{\frac12(\alpha+\beta+1)+n}}{1+q^{\frac12(\alpha+\beta+1)}},
  \ \ \check{P}_n(x_3;\bm{\lambda})=\frac{(-1)^nq^{\frac12(\alpha-\beta)n}
  (q^{\beta+1};q)_n}{(q;q)_n},\\
  &D_n^{(0,1,2,3)}(\bm{\lambda})
  =-q^{\frac12\alpha-\frac52(\beta+1)}
  (1-q^{\frac12})^2(q^{\frac12\alpha}+q^{\frac12\beta})^2
  (q^{\frac12\alpha}+q^{\frac12(\beta+1)})
  (q^{\frac12(\alpha+1)}+q^{\frac12\beta})\n
  &\phantom{D_n^{(0,1,2,3)}(\bm{\lambda})=}\times
  \frac{(1-q^{\alpha+n+1})(1-q^{\beta+n+1})
  (q^{\frac12(\alpha+\beta)+n+1},q^{\frac12(\alpha+\beta)+n+2};q^{\frac12})_3}
  {(1-q^{n+1})^3(1-q^{n+2})^2(1-q^{n+3})(1+q^{\frac12(\alpha+\beta+1)+n})^2},
  \\
  &\alpha_{n,4}(\bm{\lambda})=1,\\
  &\alpha_{n,3}(\bm{\lambda})=\frac{q^{-\frac12(\beta+1)}(1+q^{\frac12})
  (q^{\frac12\alpha}-q^{\frac12\beta})(1+q^{\frac12(\alpha+\beta)+n+3})
  (1-q^{\frac12(\alpha+\beta+7)+n})}
  {(1-q^{n+4})(1-q^{\frac12(\alpha+\beta)+n+2})},\\
  &\alpha_{n,2}(\bm{\lambda})=\frac{q^{-\beta}
  (1-q^{\frac12(\alpha+\beta+5)+n})(1-q^{\frac12(\alpha+\beta)+n+4})}
  {(q^{n+3};q)_2(q^{\frac12(\alpha+\beta+3)+n};q^{\frac12})_2}\n
  &\phantom{\alpha_{n,2}(\bm{\lambda})=}\times
  \Bigl((1+q)q^{\frac12(\alpha+\beta)-1}
  \bigl((1+q^{\frac12})^2q^{\frac12(\alpha+\beta)+n+2}
  -(1+q^{\frac12(\alpha+\beta+5)+n})^2\bigr)\n
  &\phantom{\alpha_{n,2}(\bm{\lambda})=}\qquad
  +q^{-\frac12}(q^{\frac12\alpha}-q^{\frac12\beta})^2
  \bigl((1+q)q^{\frac12(\alpha+\beta)+n+2}+(1+q^{\frac12(\alpha+\beta+5)+n})^2
  \bigr)\Bigr),\\
  &\alpha_{n,1}(\bm{\lambda})=-\frac{q^{\frac12\alpha-\beta-1}
  (1+q^{\frac12})(q^{\frac12\alpha}-q^{\frac12\beta})
  (1-q^{\alpha+n+2})(1-q^{\beta+n+2})(1+q^{\frac12(\alpha+\beta)+n+2})}
  {(q^{n+2};q)_3(q^{\frac12(\alpha+\beta)+n+1};q)_2}\n
  &\phantom{\alpha_{n,1}(\bm{\lambda})=}
  \times(q^{\frac12(\alpha+\beta+7)+n};q^{\frac12})_2,\\
  &\alpha_{n,0}(\bm{\lambda})=\frac{q^{\alpha-\beta-1}
  (q^{\alpha+n+1},q^{\beta+n+1};q)_2
  (q^{\frac12(\alpha+\beta)+n+3};q^{\frac12})_3}
  {(q^{n+1};q)_4(q^{\frac12(\alpha+\beta)+n+1};q^{\frac12})_3},\\
  &\beta_n(\bm{\lambda})=
  \frac{(q^{n+1};q)_4
  (-q^{\frac12(\alpha+\beta+1)};q^{\frac12})_4}
  {q^{\alpha-\beta-n-1}(q^{\frac12(\alpha+\beta+5)+n};q^{\frac12})_4},\quad
  \beta^{\cF}_n(\bm{\lambda})=
  \frac{q^{\beta-n}(q^{n+1};q)_4(q^{\alpha+\beta+n+3};q)_2}
  {(q^{\frac12(\alpha+\beta+5)+n};q^{\frac12})_4}.
\end{align}

\subsubsection{continuous $\bm{q}$-Laguerre (c$\bm{q}$L)}
\label{app:cqL}

\noindent
\underline{Basic data} :
\begin{align}
  &\bm{\lambda}=\alpha\ \ (\alpha\in\mathbb{R}),\quad
  \bm{\delta}=1,\\
  &V(x;\bm{\lambda})=\frac{(1-q^{\frac12(\alpha+\frac12)}e^{ix})
  (1-q^{\frac12(\alpha+\frac32)}e^{ix})}
  {(1-e^{2ix})(1-qe^{2ix})},\\
  &\mathcal{E}_n(\bm{\lambda})=q^{-n}-1,\quad
  f_n(\bm{\lambda})=
  q^{\frac12(\alpha+\frac32)}q^{-n},\quad
  b_n(\bm{\lambda})=q^{-\frac12(\alpha+\frac32)}q^{n+1}(q^{-n-1}-1),\\
  &\phi_0(x;\bm{\lambda})^2
  =\frac{(e^{2ix},e^{-2ix};q)_{\infty}}
  {(q^{\frac12(\alpha+\frac12)}e^{ix},q^{\frac12(\alpha+\frac12)}e^{-ix};
  q^{\frac12})_{\infty}},\quad
  h_n(\bm{\lambda})
  =\frac{2\pi\,q^{(\alpha+\frac12)n}(q^{\alpha+1};q)_n}
  {(q;q)_n(q,q^{\alpha+1};q)_{\infty}},\\
  &\check{P}_n(x;\bm{\lambda})=P^{(\alpha)}_n(\cos x|q)
  =\frac{(q^{\alpha+1}\,;q)_n}{(q\,;q)_n}
  {}_3\phi_2\Bigl(
  \genfrac{}{}{0pt}{}{q^{-n},\,q^{\frac12(\alpha+\frac12)}e^{ix},\,
  q^{\frac12(\alpha+\frac12)}e^{-ix}}
  {q^{\alpha+1},\,0}\Bigm|q,q\Bigr),\\
  &c_n(\bm{\lambda})=\frac{2^nq^{\frac12(\alpha+\frac12)n}}{(q;q)_n},
\end{align}
where $P^{(\alpha)}_n(\eta|q)$ is the continuous $q$-Laguerre polynomial
\cite{kls}.
The physical $\bm{\lambda}$ is $\alpha\geq-\frac12$.

\noindent
\underline{Results data} :
\begin{align}
  &\check{\Phi}(x;\bm{\lambda})
  =(q^{\frac12(\alpha+\frac12)}e^{ix},q^{\frac12(\alpha+\frac12)}e^{-ix};
  q^{\frac12})_2,\quad
  m=2,\quad c^{\Phi}(\bm{\lambda})=4q^{\alpha+1},\\
  &\eta_j=\eta(x_j),\quad
  e^{ix_j}
  =\begin{cases}
  q^{\frac12(\alpha+\frac12)}&\!\!(j=1)\\
  q^{\frac12(\alpha+\frac32)}&\!\!(j=2)
  \end{cases},\quad
  \check{P}_n(x_j;\bm{\lambda})
  =\begin{cases}
  {\displaystyle\frac{(q^{\alpha+1};q)_n}{(q;q)_n}}&\!\!(j=1)\\[8pt]
  {\displaystyle\frac{(q^{\alpha+1};q)_n}{(q;q)_n}q^{-\frac{n}{2}}}&\!\!(j=2)
  \end{cases},\\
  &D_n^{(0,1)}(\bm{\lambda})
  =\frac{(1-q^{\frac12})(1-q^{\alpha+n+1})}{q^{\frac12}(1-q^{n+1})},\\
  &\alpha_{n,2}(\bm{\lambda})=1,\quad
  \alpha_{n,1}(\bm{\lambda})=-\frac{(1+q^{\frac12})(1-q^{\alpha+n+2})}
  {q^{\frac12}(1-q^{n+2})},\quad
  \alpha_{n,0}(\bm{\lambda})=
  \frac{(q^{\alpha+n+1};q)_2}{q^{\frac12}(q^{n+1};q)_2},\\
  &\beta_n(\bm{\lambda})=q^{n+\frac12}(q^{n+1};q)_2,\quad
  \beta^{\cF}_n(\bm{\lambda})=q^{\alpha-n-\frac12}(q^{n+1};q)_2.
\end{align}

\subsection{(\romannumeral4) $\bm{\eta(x)=\cos(x+\phi)}$}
\label{app:(iv)}

\begin{align}
  &x_{\text{min}}=-\pi,\quad x_{\text{max}}=\pi,\quad\gamma=\log q,\n
  &\eta(x;\bm{\lambda})=\cos(x+\phi),\quad
  \varphi(x;\bm{\lambda})=2\sin(x+\phi),\quad\kappa=q^{-1}.
\end{align}

\subsubsection{continuous $\bm{q}$-Hahn (c$\bm{q}$H)}
\label{app:cqH}

\noindent
\underline{Basic data} :
\begin{align}
  &q^{\bm{\lambda}}=(a_1,a_2,q^{\phi})
  \ \ (a_j\in\mathbb{C},\phi\in\mathbb{R}),
  \ \ \{a_1^*,a_2^*\}=\{a_1,a_2\}\ (\text{as a set}),\quad
  \bm{\delta}=(\tfrac12,\tfrac12,0),\\
  &V(x;\bm{\lambda})=\frac{(1-a_1e^{ix})(1-a_2e^{ix})(1-a_1e^{i(x+2\phi)})
  (1-a_2e^{i(x+2\phi)})}
  {(1-e^{2i(x+\phi)})(1-qe^{2i(x+\phi)})},\\
  &\mathcal{E}_n(\bm{\lambda})=(q^{-n}-1)(1-a_1^2a_2^2q^{n-1}),
  \ \ f_n(\bm{\lambda})=q^{\frac{n}{2}}(q^{-n}-1)(1-a_1^2a_2^2q^{n-1}),
  \ \ b_n(\bm{\lambda})=q^{-\frac{n+1}{2}},\\
  &\phi_0(x;\bm{\lambda})^2
  =\frac{(e^{2i(x+\phi)},e^{-2i(x+\phi)};q)_{\infty}}
  {\prod_{j=1}^2(a_je^{ix},a_je^{-ix},a_je^{i(x+2\phi)},a_je^{-i(x+2\phi)}
  ;q)_{\infty}},\\
  &h_n(\bm{\lambda})
  =\frac{4\pi(a_1^2a_2^2q^{n-1};q)_n(a_1^2a_2^2q^{2n};q)_{\infty}}
  {(q^{n+1},a_1^2q^n,a_2^2q^n,a_1a_2q^n,a_1a_2q^n,a_1a_2e^{2i\phi}q^n,
  a_1a_2e^{-2i\phi}q^n;q)_{\infty}},\\
  &\check{P}_n(x;\bm{\lambda})=p_n\bigl(\cos(x+\phi);a_1,a_2,a_1,a_2;q\bigr)\n
  &\phantom{\check{P}_n(x;\bm{\lambda})}
  =\frac{(a_1^2,a_1a_2,a_1a_2e^{2i\phi}\,;q)_n}{a_1^ne^{in\phi}}
  {}_4\phi_3\Bigl(\genfrac{}{}{0pt}{}{q^{-n},\,a_1^2a_2^2q^{n-1},
  \,a_1e^{i(x+2\phi)},\,a_1e^{-ix}}
  {a_1^2,\,a_1a_2,\,a_1a_2e^{2i\phi}}\Bigm|q,q\Bigr),\\
  &c_n(\bm{\lambda})=2^n(a_1^2a_2^2q^{n-1};q)_n,
\end{align}
where $p_n(\eta;a_1,a_2,a_3,a_4;q)$ is the continuous $q$-Hahn polynomial
\cite{kls}.
The physical $\bm{\lambda}$ is $|a_j|<1$ ($j=1,2$).

\noindent
\underline{Results data} :
\begin{align}
  &\check{\Phi}(x;\bm{\lambda})=\prod_{j=1}^2(1-a_je^{ix})(1-a_je^{-ix})
  (1-a_je^{i(x+2\phi)})(1-a_je^{-i(x+2\phi)}),\n
  &\phantom{\check{\Phi}(x;\bm{\lambda})=}
  m=4,\quad c^{\Phi}(\bm{\lambda})=16a_1^2a_2^2,\\
  &\eta_j=\eta(x_j;\bm{\lambda}),\quad
  (e^{ix_1},e^{ix_2},e^{ix_3},e^{ix_4})
  =(a_1,a_2,a_1e^{-2i\phi},a_2e^{-2i\phi}),\\
  &\check{P}_n(x_1;\bm{\lambda})
  =\frac{(a_1^2,a_1a_2,a_1a_2e^{2i\phi};q)_n}{a_1^ne^{in\phi}},\quad
  \check{P}_n(x_2;\bm{\lambda})
  =\frac{(a_2^2,a_1a_2,a_1a_2e^{2i\phi};q)_n}{a_2^ne^{in\phi}},\n
  &\check{P}_n(x_3;\bm{\lambda})
  =\frac{(a_1^2,a_1a_2,a_1a_2e^{-2i\phi};q)_n}{a_1^ne^{-in\phi}},\quad
  \check{P}_n(x_4;\bm{\lambda})
  =\frac{(a_2^2,a_1a_2,a_1a_2e^{-2i\phi};q)_n}{a_2^ne^{-in\phi}},\\
  &D_n^{(0,1,2,3)}(\bm{\lambda})
  =4\sin^2\!\phi\,(a_1a_2)^{-5}(a_1-a_2)^2
  (a_1-a_2e^{2i\phi})(a_1-a_2e^{-2i\phi})\n
  &\phantom{D_n^{(0,1,2,3)}(\bm{\lambda})=}\times
  (1-a_1^2q^n)(1-a_2^2q^n)(1-a_1a_2q^n)^2
  (1-a_1a_2e^{2i\phi}q^n)(1-a_1a_2e^{-2i\phi}q^n)\n
  &\phantom{D_n^{(0,1,2,3)}(\bm{\lambda})=}\times
  (a_1^2a_2^2q^{2n},a_1^2a_2^2q^{2n+2};q)_3,\\
  &\alpha_{n,4}(\bm{\lambda})=1,\\
  &\alpha_{n,3}(\bm{\lambda})=
  -\frac{2\cos\phi\,(a_1+a_2)(1-a_1^2a_2^2q^{2n+5})(a_1a_2q^{n+1};q)_3}
  {a_1a_2(1-a_1^2a_2^2q^{2n+2})},\\
  &\alpha_{n,2}(\bm{\lambda})=
  \frac{(a_1^2a_2^2q^{2n+3};q^3)_2(a_1a_2q^{n+1};q)_2}
  {a_1^2a_2^2(a_1^2a_2^2q^{2n+1};q)_2}\n
  &\phantom{\alpha_{n,2}(\bm{\lambda})=}\times
  \biggl(a_1a_2(-a_1a_2q^{n+1};q)_2\bigl(2(1+a_1^2a_2^2q^{2n+3})\cos2\phi
  -(1+q)(1+q^2)a_1a_2q^n\bigr)\n
  &\phantom{\alpha_{n,2}(\bm{\lambda})=\times\Bigl(}
  +(a_1+a_2)^2\Bigl(1-(1+q)q^{n+1}(1+2\cos2\phi)a_1a_2(1+a_1^2a_2^2q^{2n+3})\\
  &\phantom{\alpha_{n,2}(\bm{\lambda})=\times\Bigl(+(a_1+a_2)^2}
  \ +a_1^2a_2^2q^{2n+2}\bigl(2q+2(1+q^2)(1+\cos 2\phi)+a_1^2a_2^2q^{2n+4}\bigr)
  \Bigr)\biggr),\n
  &\alpha_{n,1}(\bm{\lambda})=
  -\frac{2\cos\phi\,(a_1+a_2)(a_1^2a_2^2q^{2n+5};q)_2
  (a_1a_2q^n;q)_3(1-a_1a_2q^{n+1})^2}
  {a_1^2a_2^2(a_1^2a_2^2q^{2n};q^2)_2}\n
  &\phantom{\alpha_{n,1}(\bm{\lambda})=}\times
  (1-a_1^2q^{n+1})(1-a_2^2q^{n+1})
  (1-a_1a_2e^{2i\phi}q^{n+1})(1-a_1a_2e^{-2i\phi}q^{n+1}),\\
  &\alpha_{n,0}(\bm{\lambda})=
  \frac{(a_1^2a_2^2q^{2n+4};q)_3(a_1a_2q^n;q)_2^2
 (a_1^2q^n,a_2^2q^n,a_1a_2e^{2i\phi}q^n,a_1a_2e^{-2i\phi}q^n;q)_2}
  {a_1^2a_2^2(a_1^2a_2^2q^{2n};q)_3},\\
  &\beta_n(\bm{\lambda})=\frac{a_1^2a_2^2}{(a_1^2a_2^2q^{2n+3};q)_4},\quad
  \beta^{\cF}_n(\bm{\lambda})=\frac{a_1^2a_2^2q^{-n-\frac32}
  (q^{n+1},a_1^2a_2^2q^{n+1};q)_2}
  {(a_1^2a_2^2q^{2n+3};q)_4}.
\end{align}

\subsubsection{$\bm{q}$-Meixner-Pollaczek ($\bm{q}$MP)}
\label{app:qMP}

\noindent
\underline{Basic data} :
\begin{align}
  &q^{\bm{\lambda}}=(a,q^{\phi})\ \ (a,\phi\in\mathbb{R}),\quad
  \bm{\delta}=(\tfrac12,0),\\
  &V(x;\bm{\lambda})=\frac{(1-ae^{ix})(1-ae^{i(x+2\phi)})}
  {(1-e^{2i(x+\phi)})(1-qe^{2i(x+\phi)})},\\
  &\mathcal{E}_n(\bm{\lambda})=q^{-n}-1,\quad
  f_n(\bm{\lambda})=q^{-\frac{n}{2}},\quad
  b_n(\bm{\lambda})=q^{-\frac{n+1}{2}}(1-q^{n+1}),\\
  &\phi_0(x;\bm{\lambda})^2
  =\frac{(e^{2i(x+\phi)},e^{-2i(x+\phi)};q)_{\infty}}
  {(ae^{ix},ae^{-ix},ae^{i(x+2\phi)},ae^{-i(x+2\phi)};q)_{\infty}},\quad
  h_n(\bm{\lambda})
  =\frac{2\pi}{(q;q)_n(q,a^2q^n;q)_{\infty}},\\
  &\check{P}_n(x;\bm{\lambda})=p_n\bigl(\cos(x+\phi);a|q\bigr)
  =\frac{(a^2;q)_n}{a^ne^{in\phi}(q;q)_n}\,
  {}_3\phi_2\Bigl(\genfrac{}{}{0pt}{}{q^{-n},\,ae^{i(x+2\phi)},\,ae^{-ix}}
  {a^2,\,0}\Bigm|q,q\Bigr),\\
  &c_n(\bm{\lambda})=\frac{2^n}{(q;q)_n},
\end{align}
where $p_n(\eta;a|q)$ is the $q$-Meixner-Pollaczek polynomial
\cite{kls}.
The physical $\bm{\lambda}$ is $0<a<1$.

\noindent
\underline{Results data} :
\begin{align}
  &\check{\Phi}(x;\bm{\lambda})=(1-ae^{ix})(1-ae^{-ix})
  (1-ae^{i(x+2\phi)})(1-ae^{-i(x+2\phi)}),\n
  &\phantom{\check{\Phi}(x;\bm{\lambda})=}
  m=2,\quad c^{\Phi}(\bm{\lambda})=4a^2,\\
  &\eta_j=\eta(x_j;\bm{\lambda}),\quad
  (e^{ix_1},e^{ix_2})=(a,ae^{-2i\phi}),\\
  &\check{P}_n(x_1;\bm{\lambda})
  =\frac{(a^2;q)_n}{a^ne^{in\phi}(q;q)_n},\quad
  \check{P}_n(x_2;\bm{\lambda})
  =\frac{(a^2;q)_n}{a^ne^{-in\phi}(q;q)_n},\\
  &D_n^{(0,1)}(\bm{\lambda})
  =\frac{2i\sin\phi(1-a^2q^n)}{a(1-q^{n+1})},\\
  &\alpha_{n,2}(\bm{\lambda})=1,\quad
  \alpha_{n,1}(\bm{\lambda})=
  -\frac{2\cos\phi(1-a^2q^{n+1})}{a(1-q^{n+2})},\quad
  \alpha_{n,0}(\bm{\lambda})=
  \frac{(a^2q^n;q)_2}{a^2(q^{n+1};q)_2},\\
  &\beta_n(\bm{\lambda})=a^2(q^{n+1};q)_2,\quad
  \beta^{\cF}_n(\bm{\lambda})=a^2q^{-n-\frac32}(q^{n+1};q)_2.
\end{align}

\section{Explicit Expressions : oQM case}
\label{app:ExoQM}

In this appendix, data for the orthogonal polynomials described by oQM are
provided. We present basic data for the polynomials \cite{os24,fbsr}
and results data for \S\,\ref{sec:DlR}.
Data for J case is given in \S\,\ref{sec:ExJ}.

We remark that the standard parametrizations \cite{kls} are
\begin{alignat}{2}
  \text{L}&:\alpha^{\text{standard}}=g-\tfrac12,\quad&
  \text{J}&:(\alpha,\beta)^{\text{standard}}=(g-\tfrac12,h-\tfrac12),\n
  \text{B}&:a^{\text{standard}}=-2h-1,\quad&
  \text{pJ}&:(N,\nu)^{\text{standard}}=(h-\tfrac12,-\mu),
\end{alignat}
and $h$ of pJ is a continuous parameter.

\subsection{Hermite (He)}
\label{app:He}

\noindent
\underline{Basic data} :
\begin{align}
  &x_{\text{min}}=-\infty,\quad x_{\text{max}}=\infty,\quad
  \bm{\lambda}:\text{none},\\
  &\eta(x)=x,\quad\kappa=1,\quad\bm{\delta}:\text{none},\\
  &w(x;\bm{\lambda})=-\tfrac12x^2,\quad
  U(x;\bm{\lambda})=x^2-1,\\
  &c_1(\eta;\bm{\lambda})=-\tfrac12\eta,
  \ \ c_2(\eta)=\tfrac14,\ \ \constF=1,\\
  &\mathcal{E}_n(\bm{\lambda})=2n,\quad
  f_n(\bm{\lambda})=2n,\quad b_n(\bm{\lambda})=1,\\
  &\phi_0(x;\bm{\lambda})^2
  =e^{-x^2}
  =e^{-\eta(x)^2},\quad
  h_n(\bm{\lambda})
  =2^nn!\sqrt{\pi},\\
  &P_n(\eta;\bm{\lambda})
  =H_n(\eta)
  =(2\eta)^n\,{}_2F_1\Bigl(
  \genfrac{}{}{0pt}{}{-\frac{n}{2},\,-\frac{n-1}{2}}{-}
  \Bigm|-\frac{1}{\eta^2}\Bigr)
  =n!\sum_{k=0}^{[\frac{n}{2}]}\frac{(-1)^k(2\eta)^{n-2k}}{k!(n-2k)!},
  \label{cPn:H}\\
  &c_n(\bm{\lambda})=2^n,
\end{align}
where $H_n(\eta)$ is the Hermite polynomial \cite{kls} and
$[x]$ denotes the greatest integer not exceeding $x$.

\noindent
\underline{Results data} : none

\subsection{Laguerre (L)}
\label{app:L}

\noindent
\underline{Basic data} :
\begin{align}
  &x_{\text{min}}=0,\quad x_{\text{max}}=\infty,\quad
  \bm{\lambda}=g\ \ (g\in\mathbb{R}),\\
  &\eta(x)=x^2,\quad\kappa=1,\quad\bm{\delta}=1,\\
  &w(x;\bm{\lambda})=-\tfrac12x^2+g\log x,\quad
  U(x;\bm{\lambda})=x^2+\frac{g(g-1)}{x^2}-2g-1,\\
  &c_1(\eta;\bm{\lambda})=g+\tfrac12-\eta,
  \ \ c_2(\eta)=\eta,\ \ \constF=2,\\
  &\mathcal{E}_n(\bm{\lambda})=4n,\quad
  f_n(\bm{\lambda})=-2,\quad b_n(\bm{\lambda})=-2(n+1),\\
  &\phi_0(x;\bm{\lambda})^2
  =x^{2g}e^{-x^2}
  =\eta(x)^ge^{-\eta(x)},\quad
  h_n(\bm{\lambda})
  =\frac{\Gamma(n+g+\frac12)}{2\,n!},\\
  &P_n(\eta;\bm{\lambda})
  =L^{(g-\frac12)}_n(\eta)
  =\frac{(g+\frac12)_n}{n!}\,{}_1F_1\Bigl(
  \genfrac{}{}{0pt}{}{-n}{g+\frac12}\Bigm|\eta\Bigr),
  \label{cPn:L}\\
  &c_n(\bm{\lambda})=\frac{(-1)^n}{n!},
\end{align}
where $L^{(\alpha)}_n(\eta)$ is the Laguerre polynomial \cite{kls}.
The physical $\bm{\lambda}$ is $g>\frac12$.

\noindent
\underline{Results data} :
\begin{align}
  &\check{\Phi}(x)=\eta(x)=x^2,\quad
  m=1,\quad c^{\Phi}=1,\\
  &\eta_1=0,\quad
  P_n(\eta_1;\bm{\lambda})
  =\frac{(g+\frac12)_n}{n!},\\
  &D_n^{(0)}(\bm{\lambda})=1,\quad
  \alpha_{n,1}(\bm{\lambda})=1,\quad
  \alpha_{n,0}(\bm{\lambda})=-\frac{n+g+\frac12}{n+1},\\
  &\beta_n(\bm{\lambda})=-(n+1),\quad
  \beta^{\cF}_n(\bm{\lambda})=2(n+1).
\end{align}

\subsection{Bessel (B)}
\label{app:B}

\noindent
\underline{Basic data} :
\begin{align}
  &x_{\text{min}}=-\infty,\quad x_{\text{max}}=\infty,\quad
  \bm{\lambda}=h\ \ (h\in\mathbb{R}),\\
  &\eta(x)=e^x,\quad\kappa=1,\quad\bm{\delta}=-1,\\
  &w(x;\bm{\lambda})=-hx-e^{-x},\quad
  U(x;\bm{\lambda})=e^{-2x}-(2h+1)e^{-x}+h^2,\\
  &4c_1(\eta;\bm{\lambda})=2+(1-2h)\eta,
  \ \ 4c_2(\eta)=\eta^2,\ \ \constF=1,\\
  &\mathcal{E}_n(\bm{\lambda})=n(2h-n),\quad
  f_n(\bm{\lambda})=-\tfrac12n(2h-n),\quad b_n(\bm{\lambda})=-2,\\
  &\phi_0(x;\bm{\lambda})^2
  =e^{-2hx-2e^{-x}}
  =\eta(x)^{-2h}e^{-2\eta(x)^{-1}},\quad
  h_n(\bm{\lambda})
  =\frac{n!\,\Gamma(2h-n+1)}{2^{2h+1}(h-n)},\\
  &P_n(\eta;\bm{\lambda})
  =y_n(\eta;-2h-1)
  =(-1)^nn!\Bigl(\frac{\eta}{2}\Bigr)^nL^{(2h-2n)}_n(2\eta^{-1})\n
  &\phantom{P_n(\eta;\bm{\lambda})}\!
  ={}_2F_0\Bigl(
  \genfrac{}{}{0pt}{}{-n,\,n-2h}{-}\Bigm|-\frac{\eta}{2}\Bigr)
  =(n-2h)_n\Bigl(\frac{\eta}{2}\Bigr)^n{}_1F_0\Bigl(
  \genfrac{}{}{0pt}{}{-n}{2h+1-2n}\Bigm|\frac{2}{\eta}\Bigr),
  \label{cPn:B}\\
  &c_n(\bm{\lambda})=2^{-n}(n-2h)_n,
\end{align}
where $y_n(\eta;a)$ is the Bessel polynomial \cite{kls}.
The physical $\bm{\lambda}$ is $h>0$ and $\phi_n(x;\bm{\lambda})$ is
normalizable for $n=0,1,\ldots,[h]'$,
where $[x]'$ denotes the greatest integer not equal or exceeding $x$.

\noindent
\underline{Results data} :
\begin{align}
  &\check{\Phi}(x)=\eta(x)^2=e^{2x},\quad
  m=2,\quad c^{\Phi}=1,\\
  &\eta_1=0\ \ (\text{double zero}),\quad
  P_n(\eta_1;\bm{\lambda})=1,\quad
  P'_n(\eta_1;\bm{\lambda})=\tfrac12n(n-2h),\\
  &D_n^{\prime\,(0,1)}(\bm{\lambda})
  =\frac{2n-2h+1}{n(n-2h)},\\
  &\alpha_{n,2}(\bm{\lambda})=1,\quad
  \alpha_{n,1}(\bm{\lambda})=-\frac{4(n-h+1)}{2n-2h+1},\quad
  \alpha_{n,0}(\bm{\lambda})=\frac{2n-2h+3}{2n-2h+1},\\
  &\beta_n(\bm{\lambda})=\frac{4}{(2n-2h+2)_2},\quad
  \beta^{\cF}_n(\bm{\lambda})=\frac{2(n+1)(n-2h+1)}{(2n-2h+2)_2},
\end{align}
where $P'_n(\eta;\bm{\lambda})=\frac{d}{d\eta}P_n(\eta;\bm{\lambda})$.
Since $\Phi(\eta)$ has a double zero, we use Remark\,\ref{rem:Chr} of
Theorem\,\ref{thm:Chr}, and \eqref{alphank} is replaced with
\begin{equation}
  \alpha_{n,k}(\bm{\lambda})=(-1)^{2-k}
  \frac{D^{\prime\,(0,\breve{k},2)}_n(\bm{\lambda})}
  {D^{\prime\,(0,1)}_n(\bm{\lambda})},\quad
  D^{\prime\,(\ell_1,\ell_2)}_n(\bm{\lambda})
  \eqdef\left|\begin{array}{cc}
  \dfrac{P_{n+\ell_1}(\eta_1;\bm{\lambda})}{P_n(\eta_1;\bm{\lambda})}
  &\dfrac{P_{n+\ell_2}(\eta_1;\bm{\lambda})}{P_n(\eta_1;\bm{\lambda})}\\[10pt]
  \dfrac{P'_{n+\ell_1}(\eta_1;\bm{\lambda})}{P'_n(\eta_1;\bm{\lambda})}
  &\dfrac{P'_{n+\ell_2}(\eta_1;\bm{\lambda})}{P'_n(\eta_1;\bm{\lambda})}
  \end{array}\right|.
\end{equation}

\subsection{Pseudo Jacobi (pJ)}
\label{app:pJ}

\noindent
\underline{Basic data} :
\begin{align}
  &x_{\text{min}}=-\infty,\quad x_{\text{max}}=\infty,\quad
  \bm{\lambda}=(h,\mu)\ \ (h,\mu\in\mathbb{R}),\\
  &\eta(x)=\sinh x,\quad\kappa=1,\quad\bm{\delta}=(-1,0),\\
  &w(x;\bm{\lambda})=-h\log\cosh x-\mu\tan^{-1}\sinh x,\\
  &U(x;\bm{\lambda})=\frac{-h(h+1)+\mu^2+\mu(2h+1)\sinh x}{\cosh^2 x}+h^2,\\
  &4c_1(\eta;\bm{\lambda})=(1-2h)\eta-2\mu,
  \ \ 4c_2(\eta)=1+\eta^2,\ \ \constF=1,\\
  &\mathcal{E}_n(\bm{\lambda})=n(2h-n),\quad
  f_n(\bm{\lambda})=n,\quad b_n(\bm{\lambda})=2h-n-1,\\
  &\phi_0(x;\bm{\lambda})^2
  =(\cosh x)^{-2h}e^{-2\mu\tan^{-1}\sinh x}
  =\bigl(1+\eta(x)^2\bigr)^{-h}e^{-2\mu\tan^{-1}\eta(x)},\\
  &h_n(\bm{\lambda})
  =\frac{2\pi n!\,2^{2n-2h}\Gamma(2h-2n)}
  {(2h-2n+1)_n\Gamma(h-n+\frac12-i\mu)\Gamma(h-n+\frac12+i\mu)},\\
  &P_n(\eta;\bm{\lambda})
  =P_n(\eta;-\mu,h-\tfrac12)
  =\frac{(-2i)^nn!}{(n-2h)_n}P^{(-h-\frac12-i\mu,-h-\frac12+i\mu)}_n(i\eta)\n
  &\phantom{P_n(\eta;\bm{\lambda})}\!
  =\frac{(-2i)^n(-h+\frac12-i\mu)_n}{(n-2h)_n}
  {}_2F_1\Bigl(\genfrac{}{}{0pt}{}{-n,\,n-2h}{-h+\frac12-i\mu}
  \Bigm|\frac{1-i\eta}{2}\Bigr)\n
  &\phantom{P_n(\eta;\bm{\lambda})}\!
  =(\eta+i)^n{}_2F_1\Bigl(
  \genfrac{}{}{0pt}{}{-n,\,h+\frac12+i\mu-n}{2h+1-2n}
  \Bigm|\frac{2}{1-i\eta}\Bigr),
  \label{cPn:pJ}\\
  &c_n(\bm{\lambda})=1,
\end{align}
where $P_n(\eta;\nu,N)$ is the pseudo Jacobi polynomial \cite{kls}.
The physical $\bm{\lambda}$ is $h,\mu>0$ and $\phi_n(x;\bm{\lambda})$ is
normalizable for $n=0,1,\ldots,[h]'$.

\noindent
\underline{Results data} :
\begin{align}
  &\check{\Phi}(x)=1+\eta(x)^2=\cosh^2x,\quad
  m=2,\quad c^{\Phi}=1,\\
  &\eta_1=i,\quad\eta_2=-i,\quad
  P_n(\pm i\,;\bm{\lambda})=\frac{(\pm2i)^n(-h+\frac12\pm i\mu)_n}{(n-2h)_n},\\
  &D_n^{(0,1)}(\bm{\lambda})
  =-i\frac{2h-n}{h-n},\\
  &\alpha_{n,2}(\bm{\lambda})=1,\quad
  \alpha_{n,1}(\bm{\lambda})=-\frac{4\mu(2h-n-1)}{(2h-2n-1)(2h-2n-3)},\\
  &\alpha_{n,0}(\bm{\lambda})=\frac{(2h-n-1)_2\bigl((2h-2n-1)^2+4\mu^2\bigr)}
  {4(h-n-1)_2(2h-2n-1)^2},\\
  &\beta_n(\bm{\lambda})=1,\quad
  \beta^{\cF}_n(\bm{\lambda})=n+1.
\end{align}



\begin{thebibliography}{99}


\bibitem{ismail}
M.\,E.\,H.\,Ismail,
{\it Classical and Quantum Orthogonal Polynomials in One Variable\/},
vol. 98 of Encyclopedia of mathematics and its applications,
Cambridge Univ. Press, Cambridge (2005).

\bibitem{kls}
R.\,Koekoek, P.\,A.\,Lesky and R.\,F.\,Swarttouw,
{\it Hypergeometric Orthogonal Polynomials and Their $q$-Analogues,\/}
Springer-Verlag Berlin Heidelberg (2010).

\bibitem{os24}
S.\,Odake and R.\,Sasaki,
``Discrete quantum mechanics,'' (Topical Review)
J. Phys. {\bf A44} (2011) 353001 (47pp),
{\tt arXiv:1104.0473[math-ph]}.
(Typo in (2.132), $c_1(\eta,\bm{\lambda})$ for H :
$-\frac12\Rightarrow-\frac{\eta}{2}$.)

\bibitem{crum}
M.\,M.\,Crum,
``Associated Sturm-Liouville systems,''
Quart. J. Math. Oxford Ser. (2) {\bf 6} (1955) 121-127,
{\tt arXiv:physics/9908019}.
%
%
%
\bibitem{os15}
S.\,Odake and R.\,Sasaki,
``Crum's theorem for `discrete' quantum mechanics,''
Prog. Theor. Phys. {\bf 122} (2009) 1067-1079,
{\tt arXiv:0902.2593[math-ph]}.
%
%
\bibitem{os22}
S.\,Odake and R.\,Sasaki,
``Dual Christoffel transformations,''
Prog. Theor. Phys. {\bf 126} (2011) 1-34,
{\tt arXiv:1101.5468[math-ph]}.

\bibitem{is25}
M.\,E.\,H.\,Ismail and N.\,Saad,
``On the Meixner-Pollaczek polynomials and the Sturm-Liouville problems,''
J. Math. Anal. Appl. {\bf 552} (2025) 129794 (12pp),
{\tt arXiv:2507.\hspace{0pt}13583[math.CA]}.

\bibitem{os13}
S.\,Odake and R.\,Sasaki,
``Exactly solvable `discrete' quantum mechanics;
shape invariance, Heisenberg solutions,
annihilation-creation operators and coherent states,''
Prog. Theor. Phys. {\bf 119} (2008) 663-700,
{\tt arXiv:0802.1075[quant-ph]}.

\bibitem{os14}
S.\,Odake and R.\,Sasaki,
``Unified theory of exactly and quasi-exactly solvable `discrete'
quantum mechanics: I. Formalism,"
J. Math. Phys {\bf 51} (2010) 083502 (24pp),
{\tt arXiv:\hspace{0pt}0903.2604[math-ph]}.

\bibitem{jsj14}
E.I.\,Jafarov, N.I.\,Stoilova and J.\,Van der Jeugt,
``On a pair of difference equations for the ${}_4F_3$ type orthogonal
polynomials and related exactly-solvable quantum systems,''
Springer Proc. Mathematics \& Statistics {\bf 111} (2014) 291-299,
{\tt arXiv:1411.6125\hspace{0pt}[math-ph]}.

\bibitem{fbsr}
S.\,Odake,
``Another Type of Forward and Backward Shift Relations for Orthogonal
Polynomials in the Askey Scheme,''
J. Math. Anal. Appl. {\bf 540} (2024) 128591 (25pp),
{\tt arXiv:2301.00678[math.CA]}.
(For data of polynomials, see arXiv:2301.00678v1.)

\bibitem{gkm08}
D.\,G\'{o}mez-Ullate, N.\,Kamran and R.\,Milson,
%
``An extended class of orthogonal polynomials defined by a
Sturm-Liouville problem,''
J. Math. Anal. Appl. {\bf 359} (2009) 352-367,
{\tt arXiv:0807.3939[math-\hspace{0pt}ph]}.



\bibitem{os25}
S.\,Odake and R.\,Sasaki,
``Exactly solvable quantum mechanics and infinite families of
multi-indexed orthogonal polynomials,''
Phys. Lett. {\bf B702} (2011) 164-170,
{\tt arXiv:1105.\hspace{0pt}0508[math-ph]}.
(Remark: $\tilde{\bm{\delta}}^{\I}$ and $\tilde{\bm{\delta}}^{\II}$ in this
paper are changed to $-\tilde{\bm{\delta}}^{\I}$ and
$-\tilde{\bm{\delta}}^{\II}$ in the later references.)

\bibitem{os26}
S.\,Odake and R.\,Sasaki,
``Multi-indexed ($q$-)Racah polynomials,''
J. Phys. {\bf A 45} (2012) 385201 (21pp),
{\tt arXiv:1203.5868[math-ph]}.

\bibitem{os27}
S.\,Odake and R.\,Sasaki,
``Multi-indexed Wilson and Askey-Wilson polynomials,''
J. Phys. {\bf A46} (2013) 045204 (22pp),
{\tt arXiv:1207.5584[math-ph]}.


\bibitem{idQMcH}
S.\,Odake,
``Exactly Solvable Discrete Quantum Mechanical Systems and Multi-indexed
Orthogonal Polynomials of the Continuous Hahn and Meixner-Pollaczek Types,''
Prog. Theor. Exp. Phys. {\bf 2019} (2019) 123A01 (20pp),
{\tt arXiv:1907.12218[math-ph]}.


\end{thebibliography}
\end{document}